%% file: paper.tex
\documentclass{mathincs}
\input{00_customisation.tex}

\input{tikz_macros.tex}

 \newtheorem{lem}{Lemma}
\begin{document}

\title[Encapsulating Formal Methods in DSLs]{Encapsulating Formal Methods within
  Domain Specific Languages: \\ A Solution for Verifying Railway Scheme Plans}

\author{Phillip James}
\address{Department of Computer Science,\\Swansea University,\\ Singleton Park, \\Swansea,\\SA2 8PP.}
\email{cspj@swansea.ac.uk}

\author{Markus Roggenbach}
\address{Department of Computer Science,\\Swansea University,\\ Singleton Park, \\Swansea,\\SA2 8PP.}
\email{m.roggenbach@swansea.ac.uk}

\begin{abstract}
  The development and application of formal methods is a long standing
  research topic within the field of computer science. One particular
  challenge that remains is the uptake of formal methods into
  industrial practices. This paper introduces a methodology for
  developing domain specific languages for modelling and verification
  to aid in the uptake of formal methods within industry. It
  illustrates the successful application of this methodology within
  the railway domain. The presented methodology addresses issues
  surrounding faithful modelling, scalability of verification and
  accessibility to modelling and verification processes for
  practitioners within the domain.
\end{abstract}



\maketitle

\vspace{-1.3cm}
\section{Introduction}
\input{1_introduction}

\section{A Methodology for Encapsulating Formal Methods within DSLs} 
\label{sec:methodology}
\input{2_method_intro.tex}
\input{21_dsls.tex}
\input{22_design_process.tex}
\input{23_verification_process.tex}

\section{The Railway Signalling Domain}
\label{sec:railway}
\input{3_railway_intro.tex}
\input{31_railway.tex}
\input{32_bjoerner.tex}

\section{Formalizing the Signalling Domain}
\label{sec:formalization}
\input{4_formalizing_intro.tex}
\input{41_CaslAndModalCasl.tex}
\input{42_comorphism.tex}
\input{43_narrative_modelling.tex}

\section{Automating Verification for the Signalling Domain}
\label{sec:verification}
\input{5_verification_intro.tex}
\input{51_theorem_proving.tex}
\input{52_unsuccessful_verification.tex}
\input{53_supporting_verification.tex}

\section{Encapsulation in a Graphical Tooling Environment}
\label{sec:tooling}
\input{6_graphical_intro.tex}
\input{61_gmf_epsilon.tex}
\input{62_ontrack.tex}
\input{63_verification.tex}

\section{Related Work}
\input{7_relatedWork.tex}

\section{Reflection on Industrial Collaboration}
\input{8_reflection.tex}

\section{Summary \& Conclusion}
\input{9_conclusion.tex}

\bibliographystyle{abbrv}
\bibliography{doc}

\pagebreak

\appendix

\input{appendix_proof.tex}

\input{appendix_spec_structure.tex}

\end{document}

%% file: 00_customisation.tex
\usepackage{microtype}
\usepackage{fancybox}
\usepackage{float}
\usepackage{hetcasl}
\usepackage{amsmath,amssymb,amsthm,stmaryrd,mathrsfs}
\usepackage{subfloat}
\usepackage{rotating}
\usepackage{caption}
\usepackage{framed}
\usepackage{subfigure}
\usepackage{url}

\renewcommand{\tt}{\texttt}
\renewcommand{\it}{\textit}
\renewcommand{\bf}{\textbf}

\usepackage{todonotes}

\newcommand{\CASL}{\textrm{\textsc{Casl}}\xspace}

\newcommand{\ModalCASL}{\textrm{\textsc{ModalCasl}}\xspace }

\newcommand{\DB}{Dines Bj{\o}rner\xspace}
\newcommand{\Bjoerner}{Bj{\o}rner\xspace}

\newcommand{\CSPB}{CSP$||$B\xspace}

\newcommand{\Hets}{\textrm{\textsc{Hets}}\xspace}
\newcommand{\SPASS}{SPASS\xspace}

\usepackage{listings}

%% file: tikz_macros.tex
\usepackage{tikz}
\usetikzlibrary{arrows}
\usetikzlibrary{automata}
\usetikzlibrary{backgrounds}
\usetikzlibrary{calc}
\usetikzlibrary{fit}
\usetikzlibrary{shapes}
\usetikzlibrary{snakes}

\tikzset{sortTZ/.style ={}}
\tikzset{subsortTZ/.style = {draw, <-}}
\tikzset{isomorphicTZ/.style = {subsortTZ, <->}}
\tikzset{fixedTextHeightProblem/.style = {text height=1.5ex,text depth=.25ex}}

\newcommand{\RWConnectorHalfHeight}{1mm}
\newcommand{\RWPointHeight}{10mm}
\newcommand{\RWPointHalfWidth}{10mm}

\newcommand{\RWLongPointHalfWidth}{30mm}

\newcommand{\RWConnector}[2]{
  \coordinate (#1) at #2;
  \draw ($(#1) +(0,-\RWConnectorHalfHeight)$) -- ($(#1) +(0,+\RWConnectorHalfHeight)$);
}

\newcommand{\RWLabelConnectorBelow}[2]{
  \node [anchor = north] at ($(#1) +(0,-\RWConnectorHalfHeight)$) {#2};
}

\newcommand{\RWLabelLinearUnitAbove}[3]{
  \node [anchor = south] at ($(#1)!.5!(#2) +(0,+\RWConnectorHalfHeight)$) {#3};
}

\newcommand{\RWLabelLinearUnitBelow}[3]{
  \node [anchor = north] at ($(#1)!.5!(#2) +(0,-\RWConnectorHalfHeight)$) {#3};
}

\newcommand{\RWLinearUnitAbove}[3] {
  \draw (#1) -- (#2);
  \RWLabelLinearUnitAbove{#1}{#2}{#3}
}
\newcommand{\RWLinearUnitBelow}[3] {
  \draw (#1) -- (#2);
  \RWLabelLinearUnitBelow{#1}{#2}{#3}
}

\newcommand{\RWPoint}[5]{
  \coordinate (#1) at #5;
  \RWConnector{#2}{($#5 + (-\RWPointHalfWidth,0)$)}
  \RWConnector{#3}{($#5 + (\RWPointHalfWidth,0)$)}
  \RWConnector{#4}{($#5 + (\RWPointHalfWidth,\RWPointHeight)$)}

  \draw (#2) -- #5;
  \draw #5 -- (#3);
  \draw #5 -- (#4);
}

\newcommand{\RWPointUpsideDown}[5]{
  \coordinate (#1) at #5;
  \RWConnector{#2}{($#5 + (-\RWPointHalfWidth,0)$)}
  \RWConnector{#3}{($#5 + (\RWPointHalfWidth,0)$)}
  \RWConnector{#4}{($#5 + (\RWPointHalfWidth,-\RWPointHeight)$)}

  \draw (#2) -- #5;
  \draw #5 -- (#3);
  \draw #5 -- (#4);
}

\newcommand{\RWPointReverse}[5]{
  \coordinate (#1) at #5;
  \RWConnector{#2}{($#5 + (+\RWPointHalfWidth,0)$)}
  \RWConnector{#3}{($#5 + (-\RWPointHalfWidth,0)$)}
  \RWConnector{#4}{($#5 + (-\RWPointHalfWidth,\RWPointHeight)$)}

  \draw (#2) -- #5;
  \draw #5 -- (#3);
  \draw #5 -- (#4);
}

\newcommand{\RWPointReverseUpsideDown}[5]{
  \coordinate (#1) at #5;
  \RWConnector{#2}{($#5 + (+\RWPointHalfWidth,0)$)}
  \RWConnector{#3}{($#5 + (-\RWPointHalfWidth,0)$)}
  \RWConnector{#4}{($#5 + (-\RWPointHalfWidth,-\RWPointHeight)$)}

  \draw (#2) -- #5;
  \draw #5 -- (#3);
  \draw #5 -- (#4);
}

\newcommand{\RWLongPoint}[5]{
  \coordinate (#1) at #5;
  \RWConnector{#2}{($#5 + (-\RWPointHalfWidth,0)$)}
  \RWConnector{#3}{($#5 + (3*\RWPointHalfWidth,0)$)}
  \RWConnector{#4}{($#5 + (\RWPointHalfWidth,\RWPointHeight)$)}

  \draw (#2) -- #5;
  \draw #5 -- (#3);
  \draw #5 -- (#4);
}

\newcommand{\RWLongPointUpsideDown}[5]{
  \coordinate (#1) at #5;
  \RWConnector{#2}{($#5 + (-\RWPointHalfWidth,0)$)}
  \RWConnector{#3}{($#5 + (3*\RWPointHalfWidth,0)$)}
  \RWConnector{#4}{($#5 + (\RWPointHalfWidth,-\RWPointHeight)$)}

  \draw (#2) -- #5;
  \draw #5 -- (#3);
  \draw #5 -- (#4);
}

\newcommand{\RWLongPointReverse}[5]{
  \coordinate (#1) at #5;
  \RWConnector{#2}{($#5 + (\RWPointHalfWidth,0)$)}
  \RWConnector{#3}{($#5 + (-\RWLongPointHalfWidth,0)$)}
  \RWConnector{#4}{($#5 + (-\RWPointHalfWidth,\RWPointHeight)$)}

  \draw (#2) -- #5;
  \draw #5 -- (#3);
  \draw #5 -- (#4);
}

\newcommand{\RWLongPointReverseUpsideDown}[5]{
  \coordinate (#1) at #5;
  \RWConnector{#2}{($#5 + (+\RWPointHalfWidth,0)$)}
  \RWConnector{#3}{($#5 + (-\RWLongPointHalfWidth,0)$)}
  \RWConnector{#4}{($#5 + (-\RWPointHalfWidth,-\RWPointHeight)$)}

  \draw (#2) -- #5;
  \draw #5 -- (#3);
  \draw #5 -- (#4);
}

%% file: 1_introduction.tex
Formal methods in software engineering have existed at least as long
as the term ``software engineering'' itself, which was coined at the
NATO Science Conference, Garmisch, 1968. In many engineering-based
application areas, such as in the railway domain, formal verification
processes have reached an impressive level of maturity as demonstrated
by various industrial case studies, e.g.\ see~\cite{boul00, winter02,
  winter03, peleska04, james10a}. Authors like Barnes also demonstrate
that formal methods can be cost effective~\cite{barnes11}.  Even
though these studies successfully illustrate the use of formal methods
from an academic perspective, adoption of formal methods within
industry is still limited~\cite{bowen06}.
From the industrial perspective, issues include
\begin{description}
\item[Faithful modelling] Do the proposed mathematical models
  faithfully represent the systems of concern? Modelling approaches
  offered from computer science are often in a form that is acceptable
  to computer scientists, but not to the engineer working within the
  domain. How can an engineer working within the domain come up with
  new models?


\item[Scalability] Does the proposed technology scale up to
  industrially sized systems in a manner that is uniformly
  applicable? Often, formal methods have been applied in a pilot to
  specific systems, but require individual, hand-crafted adaptation
  and optimisation for each new system under consideration.

\item[Accessibility] Are the methods accessible to practitioners in
  the domain of interest or is it just the developers of the approach
  who can apply them? Handling of tools for verification procedures is
  often aimed towards a computer science audience specialised in
  verification, however they are usually not manageable by engineers
  outside the field of formal methods.
\end{description}
This paper presents a new methodology addressing these three issues. The
underlying theme is that\begin{quote}
  \centering ``Domain Specific Languages (DSLs) can aid with modelling,
  verification and encapsulation of formal methods tools within a
  given domain''.
\end{quote}

We first present our methodology in general terms and then demonstrate
it on the concrete example of verifying scheme plans from the railway
domain. Our methodology is centered around the algebraic specification
language \CASL \cite{mosses04a}. \CASL provides us with a sound
semantic foundation. Furthermore, \CASL offers mature tool support for
verification. Our methodology takes as a starting point industrial
documents describing a DSL, see for instance the Invensys Rail Data
Model~\cite{DataModel}. We then stepwise develop a modelling and
automated verification process. Thanks to elements such as graphical
tooling, the process as a whole is accessible to practitioners in the
domain, verification is scalable, and models are guaranteed to be
faithful.

The paper is organised as follows: First, we discuss our methodology
in detail. Then, we introduce railway signalling as the domain in
which we demonstrate our methodology and present the DSL which will
serve as running example throughout the paper.  The next sections
apply the methodology step by step: in Section \ref{sec:formalization}
we give a method for \emph{formalising} a DSL within \CASL (Step
M1); Section \ref{sec:verification} demonstrates how to \emph{exploit
  implicit domain knowledge} for the purpose of verification (Step
M2); Section \ref{sec:tooling} presents techniques to
\emph{encapsulate formal methods} within a tooling framework (Step M3)
-- this includes the presentation of competitive verification results
for railway scheme plan verification. Finally, we place our work in
context by discussing related work.

While in the context of this paper we deal with an ``academic'' DSL,
it is worth noting that we have successfully, i.e., with the same
positive results, applied our methodology to the DSL~\cite{DataModel}
of our industrial partner -- see \cite{james14}. The chosen academic
language is of smaller extent, however, it ``covers'' all challenging
elements.

As the paper covers such diverse topics as railways, modelling in
\CASL, verification, and tooling, we present their respective
background distributed over the paper. We use the labels
``background'' and ``contribution'' to signpost the status of each
subsection. Our paper is based upon the PhD thesis of the first
author~\cite{james14} and earlier results on the topic. Complete
specifications, further details and further examples for the work we
present in this paper can be found in the thesis~\cite{james14}. In 2011, we
published a first version of our methodology~\cite{james11a}. In 2012,
we gave a first report upon the exploitation of domain knowledge for
verification~\cite{james12}. In 2013, we discussed in detail how to
formalise DSLs within \CASL~\cite{james13}. However, this paper
comprises the first complete presentation of the whole methodology.







%% file: 2_method_intro.tex
We begin by introducing the topic of domain specific languages (DSLs),
paying particular attention to the common industrial representation
used to formulate such languages in terms of UML class diagrams
accompanied by a narrative. For such industrial DSLs, we then develop
our methodology in a detailed step-by-step manner.


%% file: 21_dsls.tex
\subsection{Background: Domain Specific Languages (DSLs)}

Throughout all areas of science, one can find approaches that are
general in principle or specific to the task at hand. A general
approach gives a solution to several problems of a similar
manner. Whereas a specific approach often solves problems in a more
comprehensive manner, but can be applied to significantly less
problems. In computer science, the differences are exemplified by
general purpose languages (GPLs) and domain specific languages (DSLs)
respectively.

Domain specific languages (DSLs)~\cite{heering05}, are languages that
have been designed and tailored for a specific application
domain. DSLs aim to abstract away technical details of computer
science from the user, allowing them to create programs or
specifications without having to be an expert programmer or
specifier. Examples of DSLs include the well known Backus Naur Form or
the commonly used HTML markup language. Considering HTML, it is
designed explicitly with webpage creation in mind. It has specific
features such as \textit{elements}, \textit{tags} and
\textit{attributes} that allow the specification of structure within a
web page. The advantages of having these domain specific features are
apparent as HTML has become the de facto standard for webpage creation
thanks to its expressiveness and to its ease of application. When we
speak of expressiveness in the context of DSLs, we refer to the
question of how easy it is for a user to describe the objects they
desire. Expressiveness can often be explored by considering how
intuitive various language constructs are. This idea forms a main
theme throughout this work.

\subsubsection{DSLs Formulated using UML Class Diagrams and Narrative}

UML Class Diagrams~\cite{uml} are industrially accepted for modelling
a variety of systems across numerous domains. Often they are used to
describe all elements and relationships occurring within a domain. As
such, a UML Class Diagram can be thought of as describing a DSL. Many
tools and frameworks actually use UML class diagrams as a starting
point for the description of a DSL~\cite{gronback09,kolovos2012}. A
typical example of such an endeavour is given by the Data
Model~\cite{DataModel} of our research partner Invensys
Rail\footnote{www.invensys.com}. It aims to describe all elements
within the railway domain.

\begin{figure}
\centering
 \includegraphics[width=\linewidth]{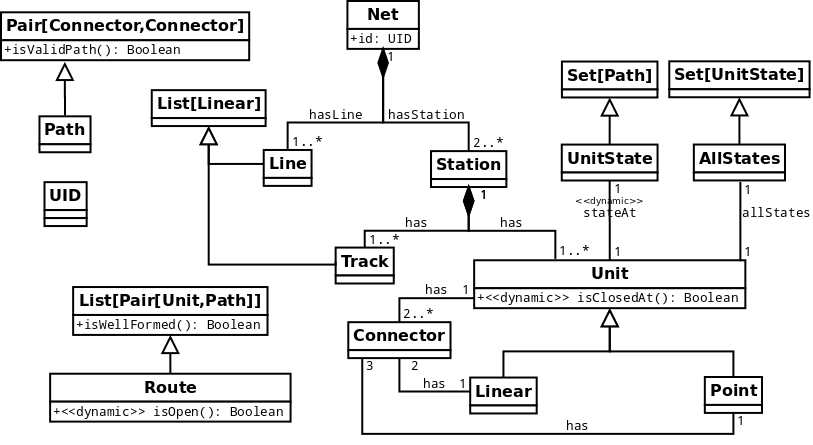}
 \caption{Part of \Bjoerner's DSL captured in a UML Diagram.}
\label{fig:bjoerners_dsl_pic}
\end{figure}

The UML diagram in Figure~\ref{fig:bjoerners_dsl_pic} captures
\Bjoerner's railway DSL~\cite{bjorner2003} that we will use as a
running example. It illustrates many of the features of class
diagrams:
\begin{itemize}
\item Classes, represented by a box, e.g.\ \emph{Net}, \emph{Unit},
  \emph{Station} etc. These represent concepts from the railway domain.
\item Properties, listed inside a class, e.g.\ \emph{id\,:\,UID} in the
  class \emph{Net} expresses that all \emph{Net}s have an identifier of
  type \emph{UID}.
\item Generalisations, represented by an unfilled arrow head, e.g.\
  \emph{Point} and \emph{Linear} are generalisations of \emph{Unit}.
\item Associations, represented by a line connecting two classes,
  e.g. the \emph{has} link between \emph{Unit} and \emph{Connector}.
  These can have direction, and also multiplicities associated with
  them. The multiplicities on the \emph{has} association between
  \emph{Unit} and \emph{Connector} can be read as: ``One \emph{Unit}
  has two or more connectors''.
\item Compositions, represented by a filled diamond, e.g. the
  \emph{hasLine} composition for \emph{Net} and \emph{Line}, tell us that
  one class ``is made up of'' another class. In a similar fashion to
  associations, compositions can also have multiplicities.
\item{Operations} are also represented inside a class, e.g.\ the
  \emph{isOpen} operation of type \emph{Boolean} inside the \emph{Route}
  class.
\end{itemize}

As UML class diagrams only capture static system aspects, we make the
realistic assumption that the class diagram is accompanied with some
narrative. Obviously, such a narrative can give explanations on the
static structure of the class diagram. Its main purpose, however, is
to describe the dynamic system aspects. For many domains such a
narrative is present in the form of standard literature. An example of
this is Kerr's narrative for the railway domain~\cite{kerr01}. Kerr
describes, for instance, how a signal changes using the following
narrative:
%
  \it{''a repeater signal shows yellow if its parent signal is
    showing red''} \cite{kerr01}.
%
Further examples of narrative are \bf{N1} and \bf{N2} in Section
\ref{ssec:industry_rw}.

UML class diagrams and narratives can be linked via the stereotype
\bf{dynamic}. This annotation indicates that the labelled class
diagram element element is related to the dynamic nature of the
system. The manner in which change happens is described in the narrative. In
Figure \ref{fig:bjoerners_dsl_pic}, e.g., the relation \it{stateAt} is
marked to be of dynamic nature, i.e., to change over time. Section
\ref{sec:railway} provides the narrative how this state change
happens.


%% file: 22_design_process.tex
\subsection{Contribution: The Design Steps of our Methodology}

To achieve a faithful, scalable and accessible modelling and
verification procedure, we present a new methodology that encapsulates
formal methods within a DSL. This results in a tool based framework
for verification.

Considering Figure~\ref{fig:process}, the encapsulation process we
propose is undertaken by a team comprising of computer scientists
working in close collaboration with experts from the domain. Here, the
close working relationship ensures the resulting domain modelling is
faithful. The following steps are involved in the process:

\begin{figure}[h]
  \centering
  \includegraphics[width=\linewidth]{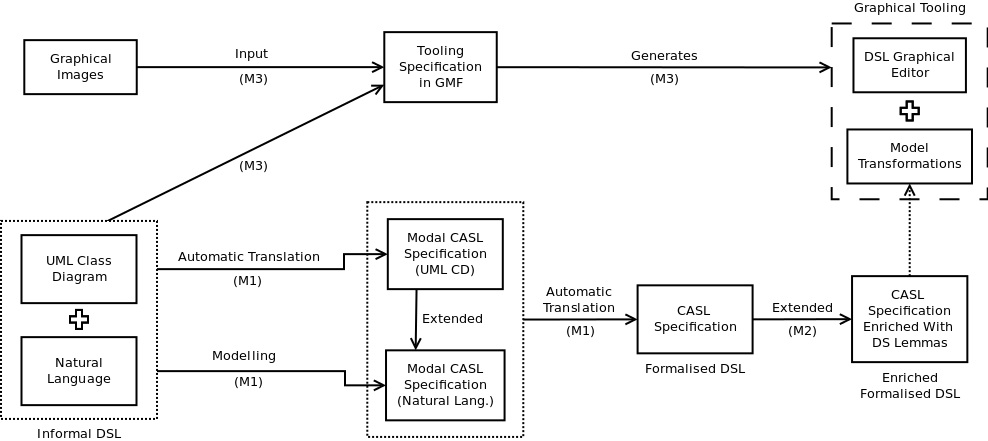}
  \caption{The proposed methodology.}
  \label{fig:process}
\end{figure}

\paragraph{M1: Formalising (Industrial) DSLs}
The starting point for our methodology is an informal domain
description in the form of UML Class Diagrams and accompanying
narrative. From such UML class diagrams, names, relations, and
multiplicity constraints can be automatically extracted and translated
into a formal specification in \ModalCASL. We suggest and support a
particular automatic translation (see
Section~\ref{sec:comorph}). Next, domain experts and computer
scientists extend the resulting formal \ModalCASL specification with a
modelling of the narrative that captures dynamics. This is possible as
\ModalCASL allows the description of state changes in terms of modal
operators. Here, we also encode proof goals for verification. Then,
for the sake of better proof support, we apply an existing automatic
translation from \ModalCASL to \CASL. Overall, starting from an
existing DSL and involving domain experts ensures a faithful
formalisation of the selected domain concepts.\\

\paragraph{M2: DSL Analysis for Verification Support}
The result of Step M1, the formalisation of the DSL, is a loose \CASL
specification -- see Section \ref{ssec:spec_languages} for
details. The logical closure of this specification, i.e., all theorems
that one can prove from this \CASL specification, is what we call the
implicitly encoded ``domain knowledge'' of the DSL. We make (part of)
this domain knowledge explicit in the form of lemmas that allow one to
refactor any proof goals into equivalent ones that are expressed on
the right level of granularity. Naturally, there cannot be a universal
solution to finding such domain specific lemmas. However, in our
experience, for all DSLs we have considered, such lemmas have existed,
follow from knowledge of the domain experts, and allow refactoring. We
discuss such lemmas in Section~\ref{sec:dsl_lemmas} and show that
these lemmas allow for scalable verification based on ideas that are
often inherent to the domain. Overall, this step enables scalability
of the verification approach.\\

\paragraph{M3: Graphical Tool Support}

DSLs are often accompanied by a development framework. For this we
make use of the \emph{Graphical Modelling Framework},
GMF~\cite{gronback09}. GMF provides the infrastructure to create, from
a UML class diagram, a Java based graphical editor. Using GMF, domain
experts and computer scientists create such a graphical tooling
environment for the DSL. This allows for native graphical
representations of domain elements. Such an editor is open (via
Epsilon~\cite{kolovos2012}) to extension with model
transformations~\cite{kolovos2012}. Such transformations allow for the
graphical models produced by the editor to be translated to \CASL
specifications. These specifications can be enriched with the domain
knowledge developed in Step M2. We illustrate this approach in
Section~\ref{sec:ontrack}, giving details of the OnTrack Toolset for
the railway domain. The result of this step is a tool for generation
of formal models that is readily usable by engineers from the domain
under consideration.

\subsubsection{Addressing the Issues}

Overall, our methodology addresses the issues we started with:
faithful modelling is achieved thanks to starting with an informal
description and forming a formal specification in a close working
relationship between the domain experts and computer scientists;
scalability of the verification procedure is achieved thanks to the
property supporting domain specific lemmas; accessibility to modelling
and verification of systems is achieved thanks to graphical tooling
incorporating domain specific concepts and constructs.


%% file: 23_verification_process.tex
\subsection{Contribution: The Resulting Verification Process}

The result of applying our methodology is a toolset accommodating the
verification process illustrated in
Figure~\ref{fig:verification_process}. This process is undertaken
purely by the engineers in the domain. It follows three main
steps:

\begin{figure}[h]
  \centering
    \includegraphics[width=\linewidth]{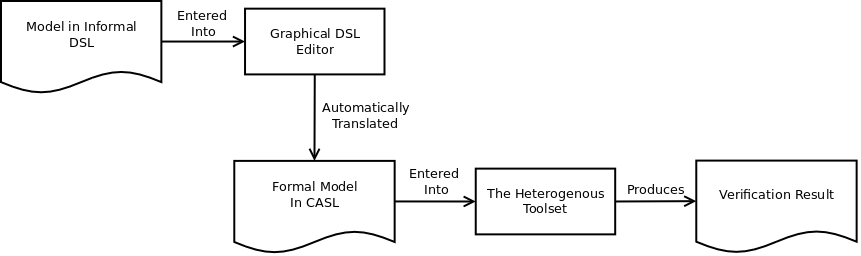}
   \caption{A verification process based on the designed tools.}
  \label{fig:verification_process}
\end{figure}

\paragraph{V1: Model Development Based on the Informal DSL} The first
(optional) step within industry is to outline or specify a design
informally. This step should be undertaken using the vocabulary
outlined within the informal DSL.\\

\paragraph{V2: Graphical Modelling} Next, the domain engineer can
encode their design using the graphical editor. Once encoded, the
engineer can automatically produce formal specifications ready for
verification. As the graphical editor contains constructs that are
from the informal DSL, training and learning costs are minimal.\\

\paragraph{V3: Verification} Finally, the formal specifications can be
verified using (for \CASL) the Heterogeneous Toolset,
\Hets~\cite{hets07}. Due to the domain specific lemmas developed
during the design process, verification is automated and ``Push
Button''.


%% file: 3_railway_intro.tex
We now introduce the railway signalling domain. We discuss the main
elements of this domain and present an established DSL by
\Bjoerner~\cite{bjorner2003} that captures it. Here, we first present
the narrative before discussing the UML class diagram.


%% file: 31_railway.tex
\subsection{Background: Industrial Practice in the Railway Domain}\label{ssec:industry_rw}

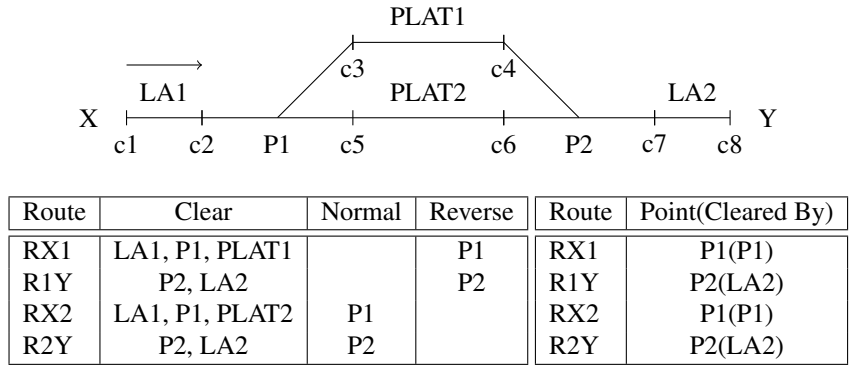
\begin{figure}[h]
  \centering
\begin{tikzpicture}[transform shape]
 \node at (-0.5,0) {X};
 \node at (8.5,0) {Y};
 \draw [->] (0,0.7) to (1,0.7);
 \RWConnector{a}{(0,0)}
 \RWLabelConnectorBelow{a}{c1}
 \RWConnector{a1}{(1,0)}
 \RWLabelConnectorBelow{a1}{c2}
 \RWLinearUnitAbove{a}{a1}{LA1}
 \RWPoint{p1}{p1cl}{p1cn}{p1cr}{(2,0)}
 \RWLabelLinearUnitBelow{p1cn}{p1cl}{P1}
 \RWLabelConnectorBelow{p1cn}{c5}
 \RWConnector{a3}{(5,0)}
 \RWLabelConnectorBelow{a3}{c6}
 \RWLinearUnitAbove{p1cn}{a3}{PLAT2}
 \RWPointReverse{p2}{p2cl}{p2cn}{p2cr}{(6,0)}
 \RWLabelLinearUnitBelow{p2cn}{p2cl}{P2}
 \RWConnector{b1}{(3,1)}
 \RWLabelConnectorBelow{b1}{c3}
 \RWConnector{b2}{(5,1)}
 \RWLabelConnectorBelow{b2}{c4}
 \RWLinearUnitAbove{b1}{b2}{PLAT1}
 \RWConnector{b4}{(7,0)}
 \RWLabelConnectorBelow{b4}{c7}
 \RWConnector{b5}{(8,0)}
  \RWLabelConnectorBelow{b5}{c8}
 \RWLinearUnitAbove{b4}{b5}{LA2}
 \end{tikzpicture}              
\\ \quad \\
\begin{tabular}{|l|c|c|c|}
\hline
Route & Clear        & Normal & Reverse \\ \hline \hline
RX1   & LA1, P1, PLAT1 &      & P1        \\
R1Y   & P2, LA2      &     & P2        \\
RX2   & LA1, P1, PLAT2 & P1       &       \\
R2Y   & P2, LA2      & P2       &       \\ \hline
\end{tabular}
\begin{tabular}{|l|c|}
\hline
Route & Point(Cleared By)               \\ \hline \hline
RX1   & P1(P1)                         \\ 
R1Y   & P2(LA2)                         \\ 
RX2   & P1(P1)                         \\ 
R2Y   & P2(LA2)                         \\ \hline 
\end{tabular}\\ \quad \\
\caption{A scheme plan for a simple station. Top: Track
  plan, Bottom Left: Control table, Bottom Right: Release table.}
\label{fig:trackplan}
\end{figure}

In industry, companies such as our industrial partner, Invensys Rail,
undertake domain engineering with the aim to ``describe all concepts,
components and properties within the railway
domain''~\cite{DataModel}. This modelling, for example, includes
features such as rail topology (the basic graph underlying the
railway), dimensions (e.g, where tracks are with regards to reference
points, the length of tracks, etc), and signalling (routes, speed
restrictions etc.). Common across all these layers is the notion of a
\emph{track plan} as a term to describe layouts of junctions and
stations. Track plans combine topological information and the
conceptual abstraction of routes, which determine the use of a rail
layout.
An example track plan is shown in Figure~\ref{fig:trackplan}. 

The intended operation of the train station shown in
Figure~\ref{fig:trackplan} is: (1) trains enter at X using track LA1,
they then proceed across point P1 towards the upper line to platform
PLAT1 (i.e.\ taking route RX1); (2) alternatively, they pass over
point P1 towards the lower line and proceed to platform PLAT2 (i.e.\
taking route RX2); (3) trains from platform PLAT1 can then pass back
the lower line using point P2 and exit the station through track LA2
(i.e.\ taking route R1Y); (4) finally, trains from platform PLAT2 can
pass across point P2 and exit the station through track LA2 (i.e.\
taking route R2Y).

Such a track plan is usually paired with a set of control and release
tables~\cite{kerr01} to form a scheme plan -- see Figure
\ref{fig:trackplan}. A scheme plan details the conditions required for
route availability. The operational setting and unsetting of points
and routes is controlled by an interlocking which is implemented based
on this scheme plan.

The control table prescribes that a given route can be used when all
tracks in the ``clear'' column are not occupied by a train, and the
points in the ``normal'' and ``reverse'' columns are set to those
positions. The example control table prescribes that route RX1 can be
assigned to a train when units LA1, P1, and PLAT1 are unoccupied and
point P1 is in its reverse position, that is, allowing trains to
travel to the top line of Figure~\ref{fig:trackplan}. We note that the
track plan in Figure~\ref{fig:trackplan} is uni-directional and that
if it were bi-directional, there would be routes corresponding to the
opposite direction of travel. The rules for these routes would also
share tracks with the current rules, stopping the possibility of
routes in opposite directions being used at the same time.

The interlocking also allocates so called locks on points to
particular route requests. These locks ensure that the point remains
locked in position. Such locks are then released according to the
information in the release table. For example, the first row of the
release table states that for route RX1 the point P1 can be released
by the point itself becoming clear. Releasing of these locks allows
the corresponding points to be used within another route. Notice, that
for a point which splits two routes, i.e.\ point P1, the lock on the
point can be released by the point itself. However, for points that
merge two routes, i.e.\ point P2, there is the added safety check that
the point can only be released after the shared parts of the routes
are cleared, i.e.\ track LA2. 


Finally, the last element in the dynamic operation of railways is that
of train movements. Above, we have described how access to certain
routes is granted, and areas of tracks can be released. This follows
conventional railway signalling~\cite{kerr01}. However, the newer
ETCS~\cite{etcs} standard builds on this conventional signalling with
the notion of a movement authority. A movement authority can be
thought of as an area of railway for which a train can be granted
access to travel along. For example, a train may be granted access to
move along units LA1 and P1 in Figure~\ref{fig:trackplan}. The
assignment of movement authorities is given by the following
narratives:
\begin{description}
\item[N1 -- Extension] Initially, no train has a movement
  authority. A request can be made for a train to travel along a
  particular route. If the route is available (as dictated by the
  control table) then the train's movement authority can be extended to
  include the route.  The movement authority for a train can contain
  multiple routes. 
\item[N2 -- Release] As a train travels it releases regions of railway
  from its movement authority according to the release table for that
  route.
\end{description}
Such movement authorities allow the example run illustrated in
Figure~\ref{fig:example_movements}. In the beginning, at time $0$,
there is no train in the system. At time $1$, train A has been
detected by track circuit $LA1$. Train A travels to platform PLAT1
where it resides until it has been overtaken by train B. Train A then
travels further and leaves the system. This run illustrates that train
A releases the use of point P1 before it exits the full route
RX1. This allows for train B to use route RX2 whilst the end of route
RX1 is still in use by train A.

\begin{figure}[h]
  \centering
\begin{footnotesize}
\begin{tabular}{l|c|c|c|c|c|c}
Time & LA1 & P1 & PLAT1 & PLAT2 & P2 & LA2 \\ \hline \hline
0 & \_ & \_ & \_ & \_ & \_ & \_ \ \\
1 & $\Longmapsto_a$ & \_ & \_ & \_ & \_ & \_ \\
2 & \_ & $\Longmapsto_a$ & \_ & \_ & \_ & \_ \\
3 & \ & \_ & $\Longmapsto_a$ & \_ & \_ & \_ \\
4 & $\Longmapsto_b$ & \_ & $\Longmapsto_a$ & \_ & \_ & \_ \\
5 & $\_$ & $\Longmapsto_b$ & $\Longmapsto_a$ & \_ & \_ & \_ \\
6 & $\_$ & $\_$ & $\Longmapsto_a$ & $\Longmapsto_b$ & \_ & \_ \\
7 & $\_$ & $\_$ & $\Longmapsto_a$ & \_ & $\Longmapsto_b$ & \_ \\
8 & $\_$ & $\_$ & $\Longmapsto_a$ & \_ & \_ & $\Longmapsto_b$ \\
9 & $\_$ & $\_$ & $\Longmapsto_a$ & \_ & \_ & \_ \\
10 & $\_$ & $\_$ & \_ & \_ &  $\Longmapsto_a$ & \_ \\
11 & $\_$ & $\_$ & \_ & \_ & \_ & $\Longmapsto_a$ \\
12 & \_ & \_ & \_ & \_ & \_ & \_ \ \\
\end{tabular}
\end{footnotesize}
\caption{A time/position diagram for an example run of the station.}
  \label{fig:example_movements}
\end{figure}

\subsubsection{Discussion of Safety Properties}
\label{sec:safety_discussion}

In railway signalling, many different safety properties have been
considered. For example, on the concrete level of an interlocking, one
may want to check the concrete property that ``Signal x only shows
green when route y is free to use''~\cite{james10b}. Alternatively,
when checking the design of a scheme plan, Moller et al.\ check that
the control table ensures collision-freedom (excluding two trains
occupying the same track)~\cite{MNRST12HVC}. Our aim differs slightly,
as we consider the assignment of movement authorities which is
outlined by the ETCS standard~\cite{etcs}. Therefore we verify that
 ``overlapping movement authorities are not assigned at the same
  time''.

Such a property is at a higher level of abstraction than the
properties mentioned previously. However, as movement authorities are
extended depending on the rules of the control table, we do in fact
cover the property of collision freedom under the assumption that
trains are well behaved. That is, if trains stay within their given
movement authority, and movement authorities are proven to never
overlap, then we know two trains cannot occupy the same track unit.


%% file: 32_bjoerner.tex
\subsection{Background: \Bjoerner's DSL Adapted for ETCS}

The process of identifying, classifying and precisely defining the
elements of a domain has been coined as ``Domain Engineering'' by
\DB~\cite{bjorner2009}.  \Bjoerner gives such a classification, i.e.,
DSL, for the railway domain using a narrative~\cite{bjorner2003} which
we introduce as a running example.

Figure~\ref{fig:bjoerners_dsl_pic} shows part of the UML class diagram
for \Bjoerner's DSL. A railway is a ``Net'', built from ``Station(s)''
that are connected via ``Line(s)''.  A station can have a complex
structure, including ``Tracks'', ``Switch Points'' (also called
points) and ``Linear Units''. ``Tracks'' and ``Lines'' can only
contain ``Linear Units''. All ``Unit(s)'' are attached together via
``Connector(s)''.  Along with defining these concepts, \Bjoerner
stipulates various well-formedness conditions on such a model, for
example, ``No two distinct lines and/or stations share units.'' or
``Every line of a net is connected to exactly two, distinct
stations''~\cite{bjorner2003}.

\Bjoerner's approach contains the necessary terms to describe the
track plan in Figure~\ref{fig:trackplan}. The whole track plan forms a
\emph{station}. This \emph{station} contains all elements such as
\emph{switch points} $P1$ and $P2$ along with \emph{linear units}
$LA1, PLAT1, PLAT2$, and $LA2$, and \emph{connectors} $c1,c2,\dots,c8$
for connecting these units. From this point on, we only consider the
elements of \Bjoerner's DSL that we require for modelling track plans
we are interested in. Therefore, some of the conditions stipulated by
\Bjoerner do not apply. For example, the track plan given in
Figure~\ref{fig:trackplan} is open ended. Hence, axioms regarding
closed networks, such as the condition ``all nets must contain two
stations'' (leaving no open lines), do not apply to our models.

\Bjoerner's DSL gains dynamics by attaching a state to each
unit~\cite{bjorner2003}. Each unit can be in one of several
\emph{states} at a given time. A state is represented using a set of
paths, where a path is a pair of distinct connectors $(c,c')$. A path
expresses that a train is allowed to move along a given unit from
connector $c$ to connector $c'$. To combine a unit and a path across
it, \Bjoerner introduces \emph{unit path pairs} by forming pairs from
units and paths across them. These paths allow one to describe which
direction along a unit a train is allowed to travel. Trains are not an
explicit part of \Bjoerner's DSL. Instead, \Bjoerner describes the
concept of a ``Route'', which is a dynamic ``window'' around a
train. Concretely, routes are lists of connected units and paths
through them. For example, the route from $X$ to $PLAT1$ of
Figure~\ref{fig:trackplan} would be captured as the list
$[(LA1,(c1,c2)) , (P1,(c2,c3)), (PLAT1,(c3,c4))].$ \Bjoerner
stipulates that a route can be dynamically changed over time using a
movement function that, for a given time, gives the set of assigned
routes. This movement function extends or shrinks a route by adding or
removing units at one or both of its ends. Train movements are
modelled using this function.

Finally, \Bjoerner has formalised his narrative in (the algebraic part
of) RSL~\cite{bjorner2003}. Here we adopt the \CASL specification
language rather than RSL. Our formalisation follows in great part
\Bjoerner's modelling, however, we utilise the \CASL features of
predicates, subsorting, and structuring in order to obtain a more
readable specification text. Our choice of \CASL is also due to the
greater level of proof support that is available for \CASL in the form
of the \Hets environment~\cite{hets07}, and the institutional base for
\Hets which allows us to describe how to implement a comorphism
from UML class diagrams to \CASL in
Section~\ref{sec:comorph}. Overall, the CASL models we present capture
the narrative in a similar manner to the RSL models presented by
\Bjoerner in~\cite{bjorner2003}.

%% file: 4_formalizing_intro.tex
We now present the formalisation of our DSL in \CASL. We introduce
\CASL and \ModalCASL, and discuss an automatic formalisation of UML
class diagrams. Finally, we model the railway narrative for movement
authorities.


%% file: 41_CaslAndModalCasl.tex
\subsection{Background: The Underlying Specification
  Formalisms}\label{ssec:spec_languages}
\label{sec:casl}

We introduce the relevant background on \CASL, \ModalCASL and give details on translating \ModalCASL to \CASL.

\subsubsection{CASL}

The Common Algebraic Specification Language~\cite{mosses04a}, known as
\CASL, is a specification formalism developed by the CoFI initiative
throughout the late 1990's and early 2000's in order to design a
\emph{Common Framework for Algebraic Specification and
  Development}. \CASL has \emph{basic}, \emph{structured}, and
\emph{architectural} specifications, of which we consider the first
two kinds only.

Roughly speaking, a {\CASL} \emph{basic specification} consists of a
\emph{signature} made up of sorts, operations, and predicates
(declared by means of the keywords \bf{sort}, \bf{op}, and \bf{pred},
respectively), and axioms referring to the signature items. Operations
can be partial or total. Furthermore, one may declare a \emph{subsort
  relation} on the sort symbols. Axioms are written in first-order
logic. Going one step beyond first order logic, \CASL also features
sort generation constraints for datatypes (keywords \bf{generated},
\bf{free type}).

As an example consider the \CASL specification in Figure
\ref{fig:casl_time} formalising the concept of time.  The
specification has the name \SIdIndex{Time}. It specifies the sort
symbol \Id{Time}, the constant function symbol \Ax{0}, the total
function symbol \Id{suc} from sort \Id{Time} to sort \Id{Time}, the
partial function symbol \Id{pre} from sort \Id{Time} to sort \Id{Time}
and the predicate symbol \Ax{\_\_}\Ax{<=}\Ax{\_\_} over $\Id{Time}
\times \Id{Time}$. The axiom is a first order formula stating that $0$
denotes the smallest element of sort \Id{Time}.

\begin{figure}[h]
\begin{small}
\begin{hetcasl}
  \SPEC \=\SIdIndex{Time} \Ax{=}\\ \> \SORT \IdDeclLabel{\Id{Time}}{Time}\\ \> \OPS \=\IdDeclLabel{\Ax{0}}{0} \Ax{:} \IdApplLabel{\Id{Time}}{Time};\\ \>\> \IdDeclLabel{\Id{suc}}{suc} \Ax{:} \=\IdApplLabel{\Id{Time}}{Time} \Ax{\rightarrow} \IdApplLabel{\Id{Time}}{Time};\\ \>\> \IdDeclLabel{\Id{pre}}{pre} \Ax{:} \=\IdApplLabel{\Id{Time}}{Time} \Ax{\rightarrow?} \IdApplLabel{\Id{Time}}{Time}\\ \> \PRED \=\Ax{\_\_}\Ax{<=}\Ax{\_\_} \Ax{:} \=\IdApplLabel{\Id{Time}}{Time} \Ax{\times} \IdApplLabel{\Id{Time}}{Time}\\
\>\Ax{\forall} \=\Id{n} \Ax{:}
\IdApplLabel{\Id{Time}}{Time} \Ax{\bullet} \=\IdApplLabel{\Ax{0}}{0}
\Ax{<=} \=\Id{n}\\
\KW{end}
\end{hetcasl}
\end{small}
\caption{A basic \CASL specification, formalising the concept of time.}\label{fig:casl_time}
\end{figure}

A \emph{model} of a \CASL specification is an algebra which interprets
sorts as (non empty) sets and operations and predicates as (partial)
functions and subsets respectively, in such a way that the given
axioms are satisfied. The subsorting relation is reflected by
injective coercion functions between the sets interpreting the
involved sorts (\emph{not} by subset inclusion),
see~\cite{mossakowski02,mosses04a} for further details. The collection
of all these models is called the \emph{model class} of the
specification. It is a speciality of \CASL to support \emph{loose
  specification}, i.e., to write a specification that has algebras of
different ``forms'' in its model class.

The \CASL specification \SIdIndex{Time} of Figure \ref{fig:casl_time}
has the naturals with the standard interpretations as its model, i.e.,
discrete time is a possible model. It also has the non-negative reals
as a model, i.e., dense continuous time is a possible model as
well. The \CASL specification \SIdIndex{Time} is a typical example of
a loose specification with algebras of different form: the naturals
are countable, while the non-negative reals are not countable.

Besides basic specifications, {\CASL} provides ways of building
complex (\emph{structured}) specifications out of simpler ones (the
simplest being basic specifications) by means of various
\emph{specification-building operations}. These include translation,
hiding, union, and both free and loose forms of extension. In our
presentation we make use of only a few of these, on which we elaborate
upon below.

\emph{Translations} of declared symbols to new symbols are specified
by giving lists of `maplets' of the form \(old \mapsto new\) (keyword
\bf{with}).

The signature of a \emph{union} of two specifications is the union of
their signatures. Given models over the component signatures, the
unique model over the union signature that extends each of these
models is called their \emph{amalgamation}. Clearly, not all pairs of
models over component signatures amalgamate. The models of a union
(keyword \bf {and}) are all amalgamations of the models of the
component specifications.

\emph{Extensions} (keyword \bf{then}) may specify new symbols or
merely require further properties of old ones. Extensions can be
classified by their effect on the specified model class. For instance,
an extension is called \emph{implicational} (annotation
\bf{\%implies}) if the signature and model class remain unchanged.
Note that this annotation has no effect on the semantics of a
specification: a specifier may use them to express their intentions,
tools may use them to generate proof obligations, see Section
\ref{sec:verification} for further details.

Structured specifications may be \emph{named}, and a named
specification may be \emph{generic} in the sense that it declares
\emph{parameters} that need to be \emph{instantiated} when the
specification is (re)used. Instantiation is a matter of providing an
appropriate \emph{argument specification}. 

The generic specification \SIdIndex{Pair} in Figure
\ref{fig:casl_generic} has two formal parameters, namely the
(non-named basic) specification \KW{sort} \IdDeclLabel{\Id{S}}{S} and
the (non-named basic) specification \KW{sort}
\IdDeclLabel{\Id{T}}{T}. In the specification
\SIdIndex{InstantiatedPair}, both these parameters are instantiated
with the specification
%
\KW{sort} \it{Connector}
as the
actual parameter. This results in a sort
\Id{Pair}[\IdApplLabel{\Id{Connector}}{Connector},\IdApplLabel{\Id{Connector}}{Connector}]
representing pairs over the sort \Id{Connector}. In a final step the
signature of the instantiated specification is adjusted by a renaming,
namely to rename the operation symbols \Id{first} and \Id{second} into
\Id{c1} and \Id{c2}.

\begin{figure}
\begin{small}
\begin{hetcasl}
\SPEC \=\SIdIndex{Pair} [\KW{sort} \IdDeclLabel{\Id{S}}{S}][\KW{sort} \IdDeclLabel{\Id{T}}{T}] \Ax{=}\\
\> \SORT \=\Id{Pair}[\=\IdApplLabel{\Id{S}}{S},\IdApplLabel{\Id{T}}{T}]\\
\> \OPS \=\IdDeclLabel{\Id{first}}{first} \Ax{:} \=\Id{Pair}[\=\IdApplLabel{\Id{S}}{S},\IdApplLabel{\Id{T}}{T}] \Ax{\rightarrow} \IdApplLabel{\Id{S}}{S};\\
\>\> \IdDeclLabel{\Id{second}}{second} \Ax{:} \=\Id{Pair}[\=\IdApplLabel{\Id{S}}{S},\IdApplLabel{\Id{T}}{T}] \Ax{\rightarrow} \IdApplLabel{\Id{T}}{T};\\
\>\> \IdDeclLabel{\Id{pair}}{pair} \Ax{:} \=\IdApplLabel{\Id{S}}{S} \Ax{\times} \IdApplLabel{\Id{T}}{T} \Ax{\rightarrow} \=\Id{Pair}[\=\IdApplLabel{\Id{S}}{S},\IdApplLabel{\Id{T}}{T}]\\
\> \Ax{\forall} \Id{s} \Ax{:} \IdApplLabel{\Id{S}}{S}; \=\Id{t} \Ax{:} \IdApplLabel{\Id{T}}{T} \\
\> \Ax{\bullet} \=\IdApplLabel{\Id{first}}{first}(\IdApplLabel{\Id{pair}}{pair}(\=\Id{s}, \Id{t})) \Ax{=} \Id{s}\\
\> \Ax{\bullet} \=\IdApplLabel{\Id{second}}{second}(\IdApplLabel{\Id{pair}}{pair}(\=\Id{s}, \Id{t})) \Ax{=} \Id{t}\\
\KW{end}\\

\SPEC \=\SIdIndex{InstantiatedPair} \Ax{=}\\
\>  \=\SId{Pair} \=[\KW{sort} \IdDeclLabel{\Id{Connector}}{Connector}][\KW{sort} \IdDeclLabel{\Id{Connector}}{Connector}] \\
\> \KW{with} \=\IdApplLabel{\Id{first}}{first} \Ax{:} \=\Id{Pair}[\=\IdApplLabel{\Id{Connector}}{Connector},\IdApplLabel{\Id{Connector}}{Connector}] \Ax{\rightarrow} \IdApplLabel{\Id{Connector}}{Connector} \Ax{\mapsto} \Id{c1}, \\
\>\> \IdApplLabel{\Id{second}}{second} \Ax{:} \=\Id{Pair}[\=\IdApplLabel{\Id{Connector}}{Connector},\IdApplLabel{\Id{Connector}}{Connector}] \Ax{\rightarrow} \IdApplLabel{\Id{Connector}}{Connector} \Ax{\mapsto} \Id{c2}\\
\KW{end}
\end{hetcasl}
\end{small}
\caption{A generic \CASL specification and its
  instantiation.}\label{fig:casl_generic}
\end{figure}

\subsubsection{Modal CASL}
\ModalCASL~\cite{mossakowski2004} is an extension of \CASL with
concepts from Modal Logic. We use only a small sublanguage of
\ModalCASL described below.

Roughly speaking, a {\ModalCASL} \emph{basic specification} consists
of a \emph{signature} that is a \CASL signature in which operation and
predicate symbols can be declared to have a fixed interpretation in
all worlds (keywords \KW{rigid}) or to possibly change interpretation
with regards to worlds (keyword \KW{flexible}).  Ordinary operation
and predicate declarations are treated as rigid.  \ModalCASL axioms
are \CASL axioms extended by first-order logic formulae involving the
modal operators $\Diamond$ (``there exists a reachable world such
that'') and $\Box$ (``in all reachable worlds holds'').

A \emph{model} of a \ModalCASL specification consists of a set of
worlds $W$, one of which is distinguished as the initial one, a binary
accessibility relation on $W$, and for each world $w$ a \CASL model
$M_w$, such that carrier sets and the interpretation of rigid
operation and predicate symbols are the same for all $M_w$, and the
given axioms are satisfied. Again, the collection of all these models
is called the \emph{model class} of the specification.

Finally, the \CASL \emph{specification-building operations} carry over
to \ModalCASL~\cite{mossakowski2004}.

\subsubsection{Translating from Modal CASL to CASL}
It is possible to define a semantics preserving map, mathematically
speaking a so-called institution comorphism, from \ModalCASL to
\CASL. Given the signature of the \ModalCASL specification, one adds a
sort $W$ (for ``worlds''), a constant $init : W$ and a binary
predicate $R : W \times W$ for reachability between worlds; rigid
operation and predicate symbols are kept unchanged; for flexible
operation and predicate symbols, $W$ is added as an extra
argument. Axioms are translated as expected, where $\Diamond$ and
$\Box$ are expressed via suitable quantification over $W.$


%% file: 42_comorphism.tex
\subsection{Contribution: Automatic Translation of UML Class Diagrams to CASL}
\label{sec:comorph}

The details of our automatic translation are given
in~\cite{james13}. Mathematically it is based on institution
theory. In~\cite{james13} we present a formal semantics to UML class
diagrams, extending prior work by Cengarle and
Knapp~\cite{cengarle08}. We then define a semantics preserving
translation to \ModalCASL. In this paper, we discuss the resulting
\ModalCASL and show how to formalise the narrative of our DSL.

\subsubsection{Translating \Bjoerner's DSL to Modal CASL}
We consider the result of translating the UML class diagram in
Figure~\ref{fig:bjoerners_dsl_pic} along the co-morphism defined
in~\cite{james13}.

First, classes from the class diagram are translated to \ModalCASL
sorts.  Generalisations are translated to subsorts in \ModalCASL. Note
that as part of our translation, we often include and (possibly)
instantiate specifications from the \CASL Basic Datatypes for the
built-in types from the class diagram, e.g.\ for type formers such as
$\mathsf{Pair}[c_1,c_2]$. Considering \Bjoerner's DSL we gain the
following \CASL specification for the translations of classes and
generalisations:
\begin{small}
\begin{hetcasl}
 \> {\small{}\KW{\%\%} Classes:}\\
 \> \SORTS \=\IdDeclLabel{\Id{Net}}{Net}, \IdDeclLabel{\Id{Station}}{Station}, \IdDeclLabel{\Id{Unit}}{Unit}, \ldots, \IdDeclLabel{\Id{UID}}{UID}\\
 \> {\small{}\KW{\%\%} Hierarchy:}\\
 \> \SORTS \=\IdApplLabel{\Id{Point}}{Point}, \IdApplLabel{\Id{Linear}}{Linear} \Ax{<} \IdApplLabel{\Id{Unit}}{Unit};
 \ldots;
 \=\IdApplLabel{\Id{Route}}{Route} \Ax{<} \IdApplLabel{\Id{ListPairUnitPath}}{ListPairUnitPath}
 \end{hetcasl}
\end{small}
 
Next, property declarations of the class diagram are translated to
total functions. For these functions, the classification into
``rigid'' and ``flexible'' are obtained directly from the stereotypes
in the class diagram. Thus, we obtain:
\begin{small} 
\begin{hetcasl}
 \> {\small{}\KW{\%\%} Properties:}\\
 \> \KW{rigid} \KW{op} \=\IdDeclLabel{\Id{id}}{id} \Ax{:}
 \=\IdApplLabel{\Id{Net}}{Net} \Ax{\rightarrow}?
 \IdApplLabel{\Id{UID}}{UID}\\
 \> \KW{flexible} \KW{ops}
  \IdDeclLabel{\Id{isClosedAt}}{isClosedAt} \Ax{:} \=\IdApplLabel{\Id{Unit}}{Unit} \Ax{\rightarrow?} \IdApplLabel{\Id{Boolean}}{Boolean};\\
 \>\ldots
 \end{hetcasl}
\end{small}
\ModalCASL predicates are then used to capture the composition and
association declarations. Again, rigid and dynamic elements can be
taken from the stereotypes of the class diagram. As well as these, a
predicate $\mathit{isAlive}$ is added for each sort. These predicates
are used to model ``flexible'' sort interpretations in \ModalCASL,
see~\cite{james13} for details. Considering the running example, we
obtain the following \ModalCASL:
\begin{small}
\begin{hetcasl}
 \> {\small{}\KW{\%\%} Compositions:}\\
 \> \KW{rigid} \KW{preds}
  \Ax{\_\_}\IdDeclLabel{\Id{has}}{::has:}\Ax{\_\_} \Ax{:} \=\IdApplLabel{\Id{Station}}{Station} \Ax{\times} \IdApplLabel{\Id{Unit}}{Unit};
  \Ax{\_\_}\IdDeclLabel{\Id{has}}{::has::}\Ax{\_\_}
 \Ax{:} \=\IdApplLabel{\Id{Station}}{Station} \Ax{\times}
  \IdApplLabel{\Id{Track}}{Track}; \\
\> \dots\\
  \> {\small{}\KW{\%\%} Associations: }\\
 \> \KW{rigid} \KW{preds}
  \Ax{\_\_}\IdDeclLabel{\Id{has}}{::has::}\Ax{\_\_} \Ax{:} \=\IdApplLabel{\Id{Unit}}{Unit} \Ax{\times} \IdApplLabel{\Id{Connector}}{Connector};
  \Ax{\_\_}\IdDeclLabel{\Id{has}}{::has::}\Ax{\_\_} \Ax{:}
 \=\IdApplLabel{\Id{Linear}}{Linear} \Ax{\times}
 \IdApplLabel{\Id{Connector}}{Connector}; \ldots\\
\> \dots\\
 \> {\small{}\KW{\%\%} Is Alive preds:}\\
 \> \KW{rigid} \KW{preds}
  \IdDeclLabel{\Id{isAlive}}{isAlive} \Ax{:} \IdApplLabel{\Id{Net}}{Net};
 \IdDeclLabel{\Id{isAlive}}{isAlive} \Ax{:} \IdApplLabel{\Id{Station}}{Station};
 \IdDeclLabel{\Id{isAlive}}{isAlive} \Ax{:} \IdApplLabel{\Id{Unit}}{Unit}; \ldots;
 \IdDeclLabel{\Id{isAlive}}{isAlive} \Ax{:} \IdApplLabel{\Id{UID}}{UID};
 \end{hetcasl}
\end{small}
Finally, we add axioms capturing multiplicity constraints from the
class diagram\footnote{We note that further axioms are also added for
  technical reasons regarding the \IdDeclLabel{\Id{isAlive}}{isAlive}
  predicate.}.  These can be systematically encoded in first order
logic using ``poor man's counting'', by providing the necessary number
of variables.  A typical example of the mapping for a composition
constraint from \Bjoerner's DSL is for the composition ``Station has
Unit''. Where, considering Figure~\ref{fig:bjoerners_dsl_pic} we can
see that there must be at least one Unit for every Station. The
resulting \ModalCASL axiom for this would be:

\begin{hetcasl}
\> \Ax{\bullet} \=\Ax{\forall} \Id{s} \Ax{:} \Id{Station} \Ax{\bullet} \=\Ax{\exists} \Id{u} \Ax{:} \Id{Unit} \Ax{\bullet} \IdApplLabel{\Id{has}}{has}(\=\Ax{s},\=\Ax{u}).
\end{hetcasl}

 \noindent Similarly, an example of a constraint on an association is
 the ``has'' association between Units and Connectors, where we can
 see that each Unit must have at least $2$ Connectors. The resulting
 \ModalCASL axiom for this would be:

\begin{hetcasl}
\> \Ax{\bullet} \=\Ax{\forall} \Id{u} \Ax{:} \Id{Unit} \Ax{\bullet} \=\Ax{\exists} \Id{c1}, \Id{c2} \Ax{:} \Id{Connector} \Ax{\bullet} \Ax{\neg} (\=\Id{c1}  \Ax{=} \Id{c2}) \Ax{\wedge} \IdApplLabel{\Id{has}}{has}(\=\Ax{c1},\=\Ax{u}) \Ax{\wedge} \IdApplLabel{\Id{has}}{has}(\=\Ax{c2},\=\Ax{u}).
\end{hetcasl}


\subsubsection{Translating Modal CASL to CASL}

Finally, for the sake of better proof support, we apply the existing
\ModalCASL to \CASL comorphism to gain a \CASL specification. For
example, we add the sort \Id{Time} to deal with modalities:
\begin{small}
\begin{hetcasl}
 \> \SORTS \=\IdDeclLabel{\Id{Time}}{Time}, \IdDeclLabel{\Id{Net}}{Net}, \IdDeclLabel{\Id{Station}}{Station}, \IdDeclLabel{\Id{Unit}}{Unit}, \ldots, \IdDeclLabel{\Id{UID}}{UID}
\end{hetcasl}
\end{small}
\noindent Subsort relations simply remain the same. Similarly, rigid
operations and predicates also remain the same:
\begin{small}
\begin{hetcasl}
  \> \KW{op} \=\IdDeclLabel{\Id{id}}{id} \Ax{:}
  \=\IdApplLabel{\Id{Net}}{Net} \Ax{\rightarrow}?
  \IdApplLabel{\Id{UID}}{UID}
\end{hetcasl}
\end{small}
\noindent whilst flexible operations and predicates have the sort
\Id{Time} added to their profile:
\begin{small}
\begin{hetcasl}
  \> \KW{op} \IdDeclLabel{\Id{isClosedAt}}{isClosedAt} \Ax{:}
  \=\IdApplLabel{\Id{Unit}}{Unit} \Ax{\times}
  \IdApplLabel{\Id{Time}}{Time} \Ax{\rightarrow?}
  \IdApplLabel{\Id{Boolean}}{Boolean};
 \end{hetcasl}
\end{small}
\noindent Finally, we note that as the presented examples of axioms
for multiplicity constraints are not dependent on flexible operations
or predicates, they remain unchanged.


%% file: 43_narrative_modelling.tex
\subsection{Contribution: Modelling of the Narrative in CASL}
\label{sec:narrative_model}
It is a matter of choice which specification language to use to model
the dynamic system aspects: in \ModalCASL or in \CASL. Here, for the
sake of readability, we present this modelling in \CASL. We note that,
throughout the presentation, we have specialised the sort
\IdDeclLabel{\Id{Time}}{Time} gained from applying our comorphism, to
the specification given in Figure \ref{fig:casl_time}.

\subsubsection{Introducing Regions for Movement}
\label{region_for_ma}
\begin{figure}[h]
\begin{small}
  \centering
  \begin{tikzpicture}[transform shape]
\node at (-0.5,0) {X};
\node at (8.5,0) {Y};
\draw [->] (0,0.7) to (1,0.7);
\RWConnector{a}{(0,0)}
\RWConnector{a1}{(1,0)}
\RWLinearUnitAbove{a}{a1}{LA1}
\RWPoint{p1}{p1cl}{p1cn}{p1cr}{(2,0)}
\RWLabelLinearUnitBelow{p1cn}{p1cl}{P1}
\RWConnector{a3}{(5,0)}
\RWLinearUnitAbove{p1cn}{a3}{PLAT2}
\RWPointReverse{p2}{p2cl}{p2cn}{p2cr}{(6,0)}
\RWLabelLinearUnitBelow{p2cn}{p2cl}{P2}
\RWConnector{b1}{(3,1)}
\RWConnector{b2}{(5,1)}
\RWLinearUnitAbove{b1}{b2}{PLAT1}
\RWConnector{b4}{(7,0)}
\RWConnector{b5}{(8,0)}
\RWLinearUnitAbove{b4}{b5}{LA2}

\RWConnector{R1a}{(0.2,-0.6)}
\RWConnector{R1b}{(2.8,-0.6)}
\draw [-] (0.2,-0.7) to (2.8,-0.7);
\RWLabelLinearUnitBelow{R1a}{R1b}{RG1}

\RWConnector{R2a}{(3.2,-0.6)}
\RWConnector{R2b}{(4.8,-0.6)}
\draw [-] (3.2,-0.7) to (4.8,-0.7);
\RWLabelLinearUnitBelow{R2a}{R2b}{RG2}

\RWConnector{R3a}{(3.2,1.6)}
\RWConnector{R3b}{(4.8,1.6)}
\draw [-] (3.2,1.7) to (4.8,1.7);
\RWLabelLinearUnitAbove{R3a}{R3b}{RG3}

\RWConnector{R4a}{(5.2,-0.6)}
\RWConnector{R4b}{(7.8,-0.6)}
\draw [-] (5.2,-0.7) to (7.8,-0.7);
\RWLabelLinearUnitBelow{R4a}{R4b}{RG4}
\end{tikzpicture}  
\end{small}
  \caption{Regions of the simple station.}
  \label{fig:station_regions}
\end{figure}
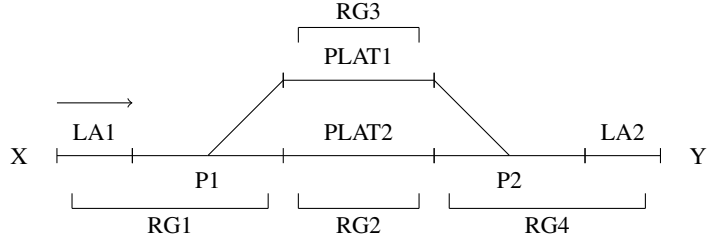

To allow us to correctly model the extension and release of movement
authorities, we introduce the notion of a region. Regions will allow
areas of a track to be reserved and released. Such regions are
categorised by both the topological routes of a track plan, and the
release points of the release table. For example, considering the pass
through station in Figure~\ref{fig:trackplan}, we can split the
topological routes at the units outlined by the release table to gain
the possible regions of the scheme plan shown in
Figure~\ref{fig:station_regions}.

The regions are defined using the release points within the release
table\footnote{We assume that a release table exhibits the property
  that all points are released ``optimally'' from a capacity point of
  view, that is, they are released as soon as the point is cleared by
  a train if travelling towards the point, or released at the end of
  the route if travelling away from a point.} to break up a route at
the given release unit. To compute the regions of a given route $r$ we
define the function $Regions:
List[Unit] \times List[Unit] \rightarrow Set[List[Unit]]$ as:\\
\\
\begin{tabular}{lll}
$Regions([u_1, \dots, u_n], [])$ & $=$ & $ \{[u_1,  \dots, u_n]\}$ \\
$Regions([u_1, \dots, u_n], [r_1, r_2, \dots, r_m])$ & $=$ & $ \{[u_1,  \dots, u_k]\} \; \cup$ \\
 & & $Regions([u_{k+1}, \dots, u_n],[r_2,\dots, r_m])$\\
& & where $u_k = r_1$.
\end{tabular}
\\ 

\noindent We then define the regions of the given route $r$ to be
$$regions(r) = Regions(units(r),releaseTable(r))$$ where $units(r)$ is the list of
units occurring within the topological route and $releaseTable(r)$ is
the topologically ordered list of release units occurring in the
release table for route $r$. For example, considering
Figure~\ref{fig:station_regions} and the route $RX1$, we have:
$Regions([LA1,P1, PLAT1],[P1]) = \{[LA1, P1]\} \; \cup \;
Regions([PLAT1],[]) = \{[LA1, P1], [PLAT1]\}.$ Notice that as we only
apply this function to routes and their release tables, each $r_i$,
for $1 \leq i \leq m,$ is guaranteed to occur only once within the
list $u_1,u_2, \dots, u_n$. That is, the region split points will be
unique for that route.

Regions allow one to model the interlocking behaviour while taking
into account the release tables. As an example, we can see that region
$RG1$ contains all the units of routes $RX1$ and $RX2$ up to and
including the release point for $P1$ as given by the release
table. Then after this release point we gain regions $RG2$ and $RG3$
representing the remaining units from the routes respectively. Such an
approach allows, the train movements given in
Figure~\ref{fig:example_movements} to be captured as the series of
region assignments as given in Figure~\ref{fig:regions}.

\begin{figure}
\centering
\begin{footnotesize}
\begin{tabular}[h]{l|c|c}
Time &  Train A Movement Authority & Train B Movement Authority \\ \hline \hline
0 & $\{ \}$ & $\{ \}$ \\
1 & $\{RG1,RG2 \}$ & $\{ \}$ \\
2 & $\{RG1,RG2 \}$ & $\{ \}$ \\
3 & $\{RG2 \}$ & $\{ \}$ \\
4 & $\{RG2 \}$ & $\{RG1,RG3 \}$ \\
5 & $\{RG2 \}$ & $\{RG1,RG3 \}$ \\
6 & $\{RG2 \}$ & $\{RG3 \}$ \\
7 & $\{RG2 \}$ & $\{RG4 \}$ \\
8 & $\{RG2 \}$ & $\{RG4 \}$ \\
9 & $\{RG2 \}$ & $\{\}$ \\
10 & $\{RG4 \}$ & $\{\}$ \\
11 & $\{RG4 \}$ & $\{\}$ \\
12 & $\{ \}$ & $\{\}$ \\
\end{tabular}
\end{footnotesize}
\caption{Modelling movement with regions.}
\label{fig:regions}
\end{figure}

To model such regions in \CASL, we instantiate a specification
\SId{List} with the sort \Id{Unit} from \Bjoerner's DSL. We then
define a subsort of these lists called \Id{Region}. Similarly, we
capture movement authorities by instantiating the specification of
\SId{List} with
these regions and making a subsort \Id{MA}:
\begin{small}
\begin{hetcasl}
\>\=\SId{List}[\KW{sort} \IdDeclLabel{\Id{Unit}}{Unit}]
\THEN \=
\SId{List}[\KW{sort} \IdDeclLabel{\Id{Region}}{Region}]
\THEN \=\\
\> \SORT \=\IdApplLabel{\Id{Region}}{Region} \Ax{<} \=\Id{List}[\IdApplLabel{\Id{Unit}}{Unit}]\\
\> \SORT \=\IdApplLabel{\Id{MA}}{MA} \Ax{<} \=\Id{List}[\IdApplLabel{\Id{Region}}{Region}]
\end{hetcasl}
\end{small}
Next, according to the rules for movement authorities given in
Section~\ref{ssec:industry_rw}, we model the assignment of a movement
authority at a given time as a predicate
\IdDeclLabel{\Id{assigned}}{assigned} \Ax{:} \IdApplLabel{\Id{MA}}{MA}
\Ax{\times} \Id{Time}. That is, if a train is assigned a movement
authority then it is allowed to travel along the railway elements
contained within the movement authority. We then add an axiom that
states that only an empty movement authority can be assigned
initially, that is at time $0$:
\begin{small}
\begin{hetcasl}
\> \PRED \=\IdDeclLabel{\Id{assigned}}{assigned} \Ax{:} \=\IdApplLabel{\Id{MA}}{MA} \Ax{\times} \Id{Time} \\
\> \Ax{\bullet} \=\Ax{\forall} \Id{m} \Ax{:} \IdApplLabel{\Id{MA}}{MA} \Ax{\bullet} \=\Ax{\neg} \=\Id{m} \Ax{=} \=\Ax{[}\Ax{]} \Ax{\Rightarrow} \Ax{\neg} \IdApplLabel{\Id{assigned}}{assigned}(\=\Id{m}, \IdApplLabel{\Ax{0}}{0}) \`{\small{}\KW{\%}(\HetsLabel{no\Ax{\_}ma\Ax{\_}0}{no:ma:0})\KW{\%}}
\end{hetcasl}
\end{small}
\noindent Similarly, to model the fact that a train can be granted a
movement authority at any time, we add an axiom that states the empty
movement authority is always available for extension:
\begin{small}
\begin{hetcasl}
\> \Ax{\bullet} \=\Ax{\forall} \Id{t} \Ax{:} \Id{Time} \Ax{\bullet} \IdApplLabel{\Id{assigned}}{assigned}(\=\Ax{[}\Ax{]} \Id{as} \IdApplLabel{\Id{MA}}{MA}, \Id{t}) 
\`{\small{}\KW{\%}(\Ax{[}\Ax{]}\_assigned\_at\_ all\_times)\KW{\%}}
\end{hetcasl}
\end{small}
\noindent We then add an axiom that states when the predicate
\IdDeclLabel{\Id{assigned}}{assigned} can hold for non empty lists. To
simplify the definition, we make use of predicates
\IdDeclLabel{\Id{canExtend}}{canExtend} \Ax{:}
\IdApplLabel{\Id{MA}}{MA} \Ax{\times} \Id{Time} and
\IdDeclLabel{\Id{canReduce}}{canReduce} \Ax{:}
\IdApplLabel{\Id{MA}}{MA} \Ax{\times} \Id{Time} whose definitions we
discuss next. The axiom given below allows for movement authorities at
some time $suc(t)$ to remain assigned as they were at time $t$, become
extended from time $t$ and also to reduce from time $t$. It also
states that only for one of these possibilities can occur for a given
movement authority at a given time. This definition matches the
expected behaviour of movement authorities given above.
\begin{small}
\begin{hetcasl}
\> \Ax{\bullet} \=\Ax{\forall} \Id{ma1} \Ax{:} \IdApplLabel{\Id{MA}}{MA}; \Id{t} \Ax{:} \Id{Time} \\
\>\> \Ax{\bullet} \=\Ax{\neg} \=\Id{ma1} \Ax{=} \=\Ax{[}\Ax{]} 
 \Ax{\Rightarrow} \=\IdApplLabel{\Id{assigned}}{assigned}(\=\Id{ma1}, \IdApplLabel{\Id{suc}}{suc}(\Id{t})) \\
\>\>\>\> \Ax{\Rightarrow} \=(\=\IdApplLabel{\Id{assigned}}{assigned}(\=\Id{ma1}, \Id{t}) \Ax{\wedge} \Ax{\neg} \IdApplLabel{\Id{canExtend}}{canExtend}(\=\Id{ma1}, \Id{t}) \Ax{\wedge} \Ax{\neg} \IdApplLabel{\Id{canReduce}}{canReduce}(\=\Id{ma1}, \Id{t})) \\
\>\>\>\>\> \Ax{\vee} (\=\Ax{\neg} \IdApplLabel{\Id{assigned}}{assigned}(\=\Id{ma1}, \Id{t}) \Ax{\wedge} \IdApplLabel{\Id{canExtend}}{canExtend}(\=\Id{ma1}, \Id{t}) \Ax{\wedge} \Ax{\neg} \IdApplLabel{\Id{canReduce}}{canReduce}(\=\Id{ma1}, \Id{t})) \\
\>\>\>\>\> \Ax{\vee} (\=\Ax{\neg} \IdApplLabel{\Id{assigned}}{assigned}(\=\Id{ma1}, \Id{t}) \Ax{\wedge} \Ax{\neg} \IdApplLabel{\Id{canExtend}}{canExtend}(\=\Id{ma1}, \Id{t}) \Ax{\wedge} \IdApplLabel{\Id{canReduce}}{canReduce}(\=\Id{ma1}, \Id{t}))\\
\`{\small{}\KW{\%}(assigned\_defn)\KW{\%}}
\end{hetcasl}
\end{small}
\noindent The predicate \IdDeclLabel{\Id{canExtend}}{canExtend} \Ax{:}
\IdApplLabel{\Id{MA}}{MA} \Ax{\times} \Id{Time} is true for a given
movement authority $ma2$, at a time $t$, when the following conditions
are met: (1) there exists a movement authority $ma1$ and a route $r$
such that the movement authority $ma1$ is assigned at $t$ and can be
extended to $ma2$ by route $r$. Here, the topological information of
valid extensions from the track plan are encoded using the predicate
\IdDeclLabel{\Id{ext}}{ext} \Ax{:} \IdApplLabel{\Id{MA}}{MA}
\Ax{\times} \IdApplLabel{\Id{Route}}{Route} \Ax{\times}
\IdApplLabel{\Id{MA}}{MA}. We control the behaviour of this predicate
by adding the following axiom:
\begin{small}
\begin{hetcasl}
\> \Ax{\bullet} \=\IdApplLabel{\Id{ext}}{ext}(\=\Id{ma1}, \Id{r}, \Id{ma2}) \Ax{\Rightarrow} \=\Id{ma2} \Ax{=} \=\Id{ma1} \Ax{++} \IdApplLabel{\Id{regions}}{regions}(\Id{r})  \`{\small{}\KW{\%}(\HetsLabel{ext\Ax{\_}defn}{ext\_defn})\KW{\%}}
\end{hetcasl}
\end{small}
\noindent stating that \IdDeclLabel{\Id{ext}}{ext} behaves like list
concatenation. (2) The route used for the extension is open at the
given time -- where the predicate
\Ax{\_\_}\IdDeclLabel{\Id{isOpenAt}}{::isOpenAt::}\Ax{\_\_} \Ax{:}
\IdApplLabel{\Id{Unit}}{Unit} \Ax{\times} \Id{Time} is used (as we
shall see later in Section~\ref{modelling_control_table}) to encode
the clear table conditions of a route; (3) If the movement authority
that is being extended is non empty\footnote{This check is required as
  we want the empty movement authority to always be assigned.}, it
becomes not assigned. This is encoded in \CASL as:
\begin{small}
\begin{hetcasl}
\> \Ax{\forall} \Id{ma1} \Ax{:} \IdApplLabel{\Id{MA}}{MA}; \=\Id{t} \Ax{:} \Id{Time} \\
\> \Ax{\bullet} \=\IdApplLabel{\Id{canExtend}}{canExtend}(\=\Id{ma1}, \Id{t}) \\
\>\> \Ax{\Leftrightarrow} \=\Ax{\exists} \Id{ma2} \Ax{:} \IdApplLabel{\Id{MA}}{MA}; \Id{r} \Ax{:} \IdApplLabel{\Id{Route}}{Route} \\
\>\>\> \Ax{\bullet} \=\IdApplLabel{\Id{assigned}}{assigned}(\=\Id{ma2}, \Id{t})  \Ax{\wedge}  \IdApplLabel{\Id{ext}}{ext}(\=\Id{ma2}, \Id{r}, \Id{ma1}) \\
\>\>\>\> \Ax{\wedge} \=\Id{r} \IdApplLabel{\Id{isOpenAt}}{::isOpenAt::} \Id{t}
 \Ax{\wedge} (\=\Ax{\neg} \=\Id{ma2} \Ax{=} \=\Ax{[}\Ax{]} \Ax{\Rightarrow} \Ax{\neg} \IdApplLabel{\Id{assigned}}{assigned}(\=\Id{ma2}, \IdApplLabel{\Id{suc}}{suc}(\Id{t}))) \`{\small{}\KW{\%}(extends\_defn)\KW{\%}}
\end{hetcasl}
\end{small}
In a similar manner, the predicate
\IdDeclLabel{\Id{canReduce}}{canReduce} \Ax{:}
\IdApplLabel{\Id{MA}}{MA} \Ax{\times} \Id{Time} is true at a given
time $t$ for a given movement authority $ma1$ if there exists a region
$rg$ such that the movement authority with the region in front of
$ma1$ was assigned at $t$ and becomes unassigned at $suc(t)$.
\begin{small}
\begin{hetcasl}
\> \Ax{\forall} \Id{ma1} \Ax{:} \IdApplLabel{\Id{MA}}{MA}; \=\Id{t} \Ax{:} \Id{Time} \\
\> \Ax{\bullet} \=\IdApplLabel{\Id{canReduce}}{canReduce}(\=\Id{ma1}, \Id{t}) \\
\>\> \Ax{\Leftrightarrow} \=\Ax{\exists} \Id{rg} \Ax{:} \IdApplLabel{\Id{Region}}{Region} \\
\>\>\> \Ax{\bullet} \=\IdApplLabel{\Id{assigned}}{assigned}(\=(\=\Id{rg} \Ax{::} \Id{ma1}) \Id{as} \IdApplLabel{\Id{MA}}{MA}, \Id{t}) \Ax{\wedge} \Ax{\neg} \IdApplLabel{\Id{assigned}}{assigned}(\=(\=\Id{rg} \Ax{::} \Id{ma1}) \Id{as} \IdApplLabel{\Id{MA}}{MA}, \IdApplLabel{\Id{suc}}{suc}(\Id{t})) \`{\small{}\KW{\%}(reduces\_defn)\KW{\%}}
\end{hetcasl}
\end{small}

Finally, we would like to ensure that only one movement authority can
be extended at any given time, this is captured using the following
axiom:
\begin{small}
\begin{hetcasl}
\> \Ax{\bullet} \=\Ax{\forall} \Id{t} \Ax{:} \Id{Time} \\
\>\> \Ax{\bullet} \=\Ax{\forall} \Id{m1}, \Id{m2} \Ax{:} \IdApplLabel{\Id{MA}}{MA} \\
\>\>\> \Ax{\bullet} \=(\=\IdApplLabel{\Id{assigned}}{assigned}(\=\Id{m1}, \IdApplLabel{\Id{suc}}{suc}(\Id{t})) 
 \Ax{\Rightarrow} \IdApplLabel{\Id{canExtend}}{canExtend}(\=\Id{m1}, \Id{t})) \Ax{\wedge}\\
\>\>\>\>\> (\=\IdApplLabel{\Id{assigned}}{assigned}(\=\Id{m2}, \IdApplLabel{\Id{suc}}{suc}(\Id{t})) \Ax{\Rightarrow} \IdApplLabel{\Id{canExtend}}{canExtend}(\=\Id{m2}, \Id{t})) \\
\>\>\>\> \Ax{\Rightarrow} \=\Id{m1} \Ax{=} \Id{m2}\`{\small{}\KW{\%}(one\_MA\_changes)\KW{\%}}
\end{hetcasl}
\end{small}
\noindent
This assumption is justified by the fact that the Invensys railway
control systems, which are responsible for one rail node, handle at
most one route request at a given time.

\subsubsection{Modelling Control Tables and Route Availability}
\label{modelling_control_table}
The above modelling of movement authorities requires the definition of
what it means for a route to be open. As we have seen, this
information is given by the control table. To model a control table,
we introduce the predicate \IdDeclLabel{\Id{clear}}{clear}~\Ax{:}
\IdApplLabel{\Id{Route}}{Route} \Ax{\times}
\IdApplLabel{\Id{Unit}}{Unit}. We then use this predicate to encode
that a particular unit occurs within the clear column for the given
route. The control table given in Figure~\ref{fig:trackplan} would be
encoded as:
\begin{small}
\begin{hetcasl}
\> \Ax{\bullet} \IdApplLabel{\Id{clear}}{clear}(\=\IdApplLabel{\Id{RX1}}{RX1}, \IdApplLabel{\Id{LA1}}{LA1})\\
\> \Ax{\bullet} \IdApplLabel{\Id{clear}}{clear}(\=\IdApplLabel{\Id{RX1}}{RX1}, \IdApplLabel{\Id{P1}}{P1})\\
\> \dots \\
\> \Ax{\bullet} \IdApplLabel{\Id{clear}}{clear}(\=\IdApplLabel{\Id{RX2}}{RX2}, \IdApplLabel{\Id{LA2}}{LA2})
\end{hetcasl}
\end{small}
\noindent In a similar manner, we could use predicates to encode the
normal and reverse columns of the control table. 
%
%
This would be a straight-forward extension of our railway
model. However, we we refrain from this as we use the railway domain
only as a proof of concept. Therefore, we assume from now on that the
normal and reverse columns in the control table are correct.
Consequently, we exclude their definitions in our modelling. Using the
above described predicates we give the following definition of what it
means for a route to be open:
\begin{small}
\begin{hetcasl}
\>  \=\Ax{\forall} \Id{r} \Ax{:} \IdApplLabel{\Id{Route}}{Route};
  \=\Id{t} \Ax{:} \Id{Time}   \Ax{\bullet} \=\Id{r} \IdApplLabel{\Id{isOpenAt}}{::isOpenAt::} \Id{t} \Ax{\Leftrightarrow} \=\Ax{\forall} \Id{u} \Ax{:} \IdApplLabel{\Id{Unit}}{Unit} \Ax{\bullet} \=\IdApplLabel{\Id{clear}}{clear}(\=\Id{r}, \Id{u}) \Ax{\Rightarrow} \=\Id{u} \IdApplLabel{\Id{isOpenAt}}{::isOpenAt::} \Id{t}
\end{hetcasl}
\end{small}
\noindent This axiom states that a route $r$ is open at a given time
$t$ if for all units $u$ for which clear($r,u$) holds, the unit $u$ is
open at $t$.

Finally, we have yet to consider what is means for a unit to be
occupied. Within the operation of an interlocking a unit is in use
when a train is detected on a particular track. In our modelling, we
abstract on the concrete position of a train and instead say that if a
region is assigned to some train, then we assume that the interlocking
knows that the region is assigned. We model this by saying that one of
the units of the region is not open for use. This is captured by the
following axiom:
\begin{small}
\begin{hetcasl}
\> \Ax{\bullet} \=\Ax{\forall} \Id{t} \Ax{:} \Id{Time}; \Id{r} \Ax{:} \Id{Route}; \Id{rg} \Ax{:} \IdApplLabel{\Id{Region}}{Region}; \Id{ma} \Ax{:} \Id{MA} \\
\>\> \Ax{\bullet}  \IdApplLabel{\Id{assigned}}{assigned}(\=\Id{ma}, \Id{t}) \Ax{\wedge} \=\Id{rg} \IdApplLabel{\Id{eps}}{::eps::} \IdApplLabel{\Id{regions}}{regions}(\Id{r}) \Ax{\wedge} \=\Id{rg} \IdApplLabel{\Id{eps}}{::eps::} \Id{ma}\\
\>\> \Ax{\Rightarrow} \=\Ax{\exists} \Id{u} \Ax{:} \IdApplLabel{\Id{Unit}}{Unit}; \Id{upp} \Ax{:} \IdApplLabel{\Id{UnitPathPair}}{UnitPathPair} \\
\>\>\> \Ax{\bullet}  \Ax{\neg} \=\Id{u} \IdApplLabel{\Id{isOpenAt}}{::isOpenAt::} \Id{t} \Ax{\wedge} \=\Id{u} \IdApplLabel{\Id{eps}}{::eps::} \Id{rg} \Ax{\wedge} \=\Id{getUnit}(\Id{upp}) \Ax{=} \Id{u} \Ax{\wedge} \=\Id{upp} \IdApplLabel{\Id{eps}}{::eps::} \Id{r} 
\`{\small{}\KW{\%}(occupied)\KW{\%}}
\end{hetcasl}
\end{small}
\noindent This axiom makes use of the operation
\IdDeclLabel{\Id{regions}}{regions} \Ax{:}
\IdApplLabel{\Id{Route}}{Route} \Ax{\rightarrow}
\IdApplLabel{\Id{MA}}{MA} that, for a given route gives the regions it
has been split into via our modelling. It also make use of the
operation \IdApplLabel{\Id{eps}}{::eps::} that simply stands for
elementhood. That is, it captures that region $r$ ``is in'' movement
authority $ma$. Overall, the axiom links the assignment of a movement
authority to the openness of a units within the movement
authority. Although this axiom is fairly loose, i.e. we do not even
impose the restriction that trains should move in the correct
direction, this is enough to prove the safety property we discuss
next.

\subsubsection{Modelling our Safety Property}
In Section~\ref{sec:safety_discussion} we discussed the safety
property that we would like to verify. To model such a property, we
introduce the auxiliary predicate \IdDeclLabel{\Id{share}}{share}
\Ax{:} \Id{MA} \Ax{\times} \Id{MA} that encodes what it means for two
movement authorities to overlap:
\begin{small}
\begin{hetcasl}
\PRED \=\IdDeclLabel{\Id{share}}{share} \Ax{:} \=\Id{MA} \Ax{\times} \Id{MA}\\
\> \Ax{\forall} \Id{ma1}, \Id{ma2} \Ax{:} \Id{MA} \\
\>\> \Ax{\bullet} \=\IdApplLabel{\Id{share}}{share}(\=\Id{ma1}, \Id{ma2}) \Ax{\Leftrightarrow} \=\Ax{\exists} \Id{rg} \Ax{:} \IdApplLabel{\Id{Region}}{Region} \Ax{\bullet} \=\Id{rg} \IdApplLabel{\Id{eps}}{::eps::} \Id{ma1} \Ax{\wedge} \=\Id{rg} \IdApplLabel{\Id{eps}}{::eps::} \Id{ma2}
\`{\small{}\KW{\%}(share\_defn)\KW{\%}}
\end{hetcasl}
\end{small}
\noindent This allows us to model safety using the following formula:
\begin{small}
\begin{hetcasl}
\=\Ax{\forall} \Id{t} \Ax{:} \Id{Time}, \Id{ma1}, \Id{ma2} \Ax{:} \Id{MA} \\
\>\> \Ax{\bullet} \=\IdApplLabel{\Id{share}}{share}(\=\Id{ma1}, \Id{ma2}) \\
\>\> \Ax{\Rightarrow} \=\Id{ma1} \Ax{=} \Id{ma2} \Ax{\vee} \Ax{\neg} (\=\IdApplLabel{\Id{assigned}}{assigned}(\=\Id{ma1}, \IdApplLabel{\Ax{t}}{t}) \Ax{\wedge} \IdApplLabel{\Id{assigned}}{assigned}(\=\Id{ma2}, \IdApplLabel{\Ax{t}}{t})) \`{\small{}\KW{\%}(safety)\KW{\%}}
\end{hetcasl}
\end{small}
\noindent It states that if two movement authorities share a region,
then either they are the same, or they are never both assigned at the
same time. This concludes our modelling for movement authorities and
safety, next we consider supporting verification of such a property.


%% file: 5_verification_intro.tex
We now consider the task of verifying our safety property over models
formulated using our formal DSL. We begin by introducing the proof
support that is available for \CASL. We then show that verification is
only possible thanks to exploitation of domain specific lemmas that we
introduce thanks to domain knowledge.


%% file: 51_theorem_proving.tex
\subsection{Background: Theorem Proving for CASL}
With respect to \CASL and \ModalCASL, tool support comes in the form
of \Hets~\cite{hets07}, the Heterogeneous Tool Set. \Hets supports
parsing (is the specification correct w.r.t.\ the context-free
grammar?) and static analysis (is the specification correct
w.r.t.\ context-sensitive properties, such as: have all sorts occurring
in a formula been declared?) of specifications written in \CASL and
\ModalCASL. \Hets also acts as a broker to various proof assistants
and automated theorem provers that can be called to discharge
proofs. We mainly make use of \SPASS~\cite{weidenbach02} and of
eProver~\cite{schulz02} both of which support theorem proving for
first-order logic with equality. At this point, we note that in our
experience with our models, both these theorem provers work best with
smaller axiomatic bases. Hence whenever possible, we make our
specifications as loose as possible by systematically excluding axioms
that are not required for a given proof.


%% file: 52_unsuccessful_verification.tex
\subsection{Contribution: Unsuccessful Verification Over \Bjoerner's
  DSL}\label{ssec:failed_verification} Overall, the capturing of
\Bjoerner's DSL along with our extension of movement authorities in
\CASL is straightforward. However, with respect to verification, both
\SPASS and eProver were unable to directly prove our safety property
(within six hours each). Even with the addition of the following
property specific axiom for induction over time, both provers still
fail:
\begin{small}
\begin{hetcasl}
\>\Ax{\bullet} (\=\Ax{\forall} \Id{ma1}, \Id{ma2} \Ax{:} \Id{MA} \Ax{\bullet} \=\IdApplLabel{\Id{share}}{share}(\=\Id{ma1}, \Id{ma2}) \\
\>\> \Ax{\Rightarrow} \=\Id{ma1} \Ax{=} \Id{ma2} \Ax{\vee} \Ax{\neg} (\=\IdApplLabel{\Id{assigned}}{assigned}(\=\Id{ma1}, \IdApplLabel{\Ax{0}}{0}) \Ax{\wedge} \IdApplLabel{\Id{assigned}}{assigned}(\=\Id{ma2}, \IdApplLabel{\Ax{0}}{0}))) \`{\small{}\KW{\%}(base case)\KW{\%}} \\ 
\>\>  \Ax{\wedge}\\
\>\> \Ax{\forall} \=\Id{t} \Ax{:} \Id{Time} \Ax{\bullet} \=(\=\Ax{\forall} \Id{ma1}, \Id{ma2} \Ax{:} \Id{MA} \Ax{\bullet} \=\IdApplLabel{\Id{share}}{share}(\=\Id{ma1}, \Id{ma2}) \\
\>\>\> \Ax{\Rightarrow} \=\Id{ma1} \Ax{=} \Id{ma2} \Ax{\vee} \Ax{\neg} (\=\IdApplLabel{\Id{assigned}}{assigned}(\=\Id{ma1}, \Id{t}) \Ax{\wedge} \IdApplLabel{\Id{assigned}}{assigned}(\=\Id{ma2}, \Id{t}))) \\
\>\> \Ax{\Rightarrow} (\=\Ax{\forall} \Id{ma1}, \Id{ma2} \Ax{:} \Id{MA} \Ax{\bullet} \=\IdApplLabel{\Id{share}}{share}(\=\Id{ma1}, \Id{ma2}) \\
\>\>\> \Ax{\Rightarrow} \=\Id{ma1} \Ax{=} \Id{ma2} \Ax{\vee} \Ax{\neg} (\=\IdApplLabel{\Id{assigned}}{assigned}(\=\Id{ma1}, \IdApplLabel{\Id{suc}}{suc}(\Id{t})) \Ax{\wedge} \IdApplLabel{\Id{assigned}}{assigned}(\=\Id{ma2}, \IdApplLabel{\Id{suc}}{suc}(\Id{t})))) \`{\small{}\KW{\%}(step case)\KW{\%}}\\ 
\> \Ax{\Rightarrow} \Ax{\forall} \=\Id{t} \Ax{:} \Id{Time} \Ax{\bullet} \=(\=\Ax{\forall} \Id{ma1}, \Id{ma2} \Ax{:} \Id{MA} \Ax{\bullet} \=\IdApplLabel{\Id{share}}{share}(\=\Id{ma1}, \Id{ma2}) \\
\>\> \Ax{\Rightarrow} \=\Id{ma1} \Ax{=} \Id{ma2} \Ax{\vee} \Ax{\neg} (\=\IdApplLabel{\Id{assigned}}{assigned}(\=\Id{ma1}, \Id{t}) \Ax{\wedge} \IdApplLabel{\Id{assigned}}{assigned}(\=\Id{ma2}, \Id{t}))) 
\end{hetcasl}
\end{small}
\noindent This is not surprising, as we, as well as \Bjoerner, have
intended to provide a strong language for modelling, and we have aimed
to model movement authorities intuitively. However, the concept of a
movement authority and our safety principle lend themselves, on the
general level of the railway domain, to natural abstraction that we
now show can be exploited for verification.


%% file: 53_supporting_verification.tex
\subsection{Contribution: Developing DSL Specific Knowledge}
\label{sec:dsl_lemmas}

Within the railway domain, it is understood that control tables are
vital in ensuring safety. In our presentation we can see that movement
authorities are extended depending on the rules of the control
table. That is, for a movement authority to be extended by the regions
of a route, that route must be open. Through domain analysis, we can
reduce the reasoning on the level of movement authorities, to a
reasoning on the level of topological routes and the control
table. This is captured by following domain specific lemma:


\pagebreak
\begin{lem}[Property Reduction to Routes]
\label{lem:dsl}
  Given a scheme plan $SP$ then
\begin{small}
\begin{hetcasl}
\> \Ax{\forall} \Id{t} \Ax{:} \Id{Time}; \Id{m1}, \Id{m2} \Ax{:} \Id{MA}
\Ax{\bullet} \IdApplLabel{\Id{share}}{share}(\Id{m1}, \Id{m2})\\
\>\> \Ax{\Rightarrow} \Id{m1} \Ax{=} \Id{m2} \Ax{\vee} \Ax{\neg} (\IdApplLabel{\Id{assigned}}{assigned}(\Id{m1}, \IdApplLabel{\Ax{t}}{t}) \Ax{\wedge} \IdApplLabel{\Id{assigned}}{assigned}(\Id{m2}, \IdApplLabel{\Ax{t}}{t})) \`{\small{}\KW{\%}(*)\KW{\%}} 
\end{hetcasl}
\end{small}
\noindent if and only if,
\begin{small}
\begin{hetcasl}
\> \Ax{\forall} \Id{t} \Ax{:} \Id{Time}; \Id{r} \Ax{:} \Id{Route}; \Id{rg} \Ax{:} \IdApplLabel{\Id{Region}}{Region}; \Id{ma} \Ax{:} \Id{MA} 
\\ 
\>\>\Ax{\bullet} \IdApplLabel{\Id{assigned}}{assigned}(\Id{ma}, \Id{t}) \Ax{\wedge} \Id{rg} \IdApplLabel{\Id{eps}}{::eps::} \Id{ma} \Ax{\wedge} \Id{rg} \IdApplLabel{\Id{eps}}{::eps::} \IdApplLabel{\Id{regions}}{regions}(\Id{r})\\
\>\>\> \Ax{\Rightarrow} \Ax{\neg} \Id{r} \IdApplLabel{\Id{isOpenAt}}{::isOpenAt::} \Id{t}  \`{\small{}\KW{\%}(**)\KW{\%}} 
\end{hetcasl}
\end{small}
\end{lem}


The proof follows by induction on time from the axioms we have
presented, including the induction axiom of Section
\ref{ssec:failed_verification}. The full proof is given in
Appendix~\ref{app:proof}. We first completed this proof by hand, and
then attempted it with \Hets. Encoding the proof for automated theorem
proving led us to consider the role of empty routes in more depth than
in the hand written proof where we had assumed certain
details. See~\cite{james14} for details on the encoding of the proof
into steps within \CASL and how these steps allow the proof to be
automatically discharged using \Hets.

The result of this lemma is that we can reduce the verification
problem over movement authorities to a simpler problem over route
openness. At this point, we note that this lemma is completely
independent of any concrete scheme plan formulated using our extended
DSL. Hence, once it has been proven as a consequence of the DSL, it
can be used to aid with verification of any scheme plan formulated
using the extended DSL. In Section~\ref{ssec:verification} we show
that this lemma is enough to make verification of large scheme plans
highly feasible. 

Finally, Appendix~\ref{app:struture} highlights, mainly for clarity,
the specification structure that we have employed throughout our
modelling.


%% file: 6_graphical_intro.tex
In this section we discuss the last step of our methodology namely the
design and implementation of graphical tool support for the developed
DSL.


%% file: 61_gmf_epsilon.tex
\subsection{Background: EMF, GMF and Epsilon}
\label{sec:eclipse-based-dsls}
Many people consider the core of a language to be its abstract
syntax. The Eclipse Modelling Framework (EMF)~\cite{steinberg08} is a
modelling framework and code generation facility for building tools
and other applications based on a structured data model. Part of this
framework includes Ecore~\cite{steinberg08} which is a UML class
diagram like language for describing meta models for DSLs. Such a
model serves as the basis for creating a graphical syntax for a DSL
using the graphical modelling framework. The Graphical Modelling
Framework (GMF)~\cite{gronback09} provides features allowing one to
develop, from an Ecore meta model, a graphical concrete syntax for a
DSL. This editor can be used to produce model instances of the DSL
described by the underlying Ecore meta model. Often, the development
of a GMF editor is motivated by the possibility of producing, from an
instance model created by the editor, some sort of output usually in
the form of text or program code. To help with this task, users can
make use of what are known as model transformations. For this task, we
make use of Epsilon~\cite{kolovos2012}. Epsilon provides a family of
languages and features for defining and applying model
transformations, comparisons, validation and code generation.



%% file: 62_ontrack.tex
\subsection{Contribution: OnTrack Editor + Model Transformations}
\label{sec:ontrack}
As we have seen, within the railway industry, defining graphical
descriptions is the de facto method of designing railway networks.  Up
until now, we have presented several \CASL models for varying aspects
of the railway domain. Although these models use terminology from the
railway domain, the modelling approaches presented are not in a form
that is common knowledge for the everyday railway engineer. In this
section, we introduce the OnTrack toolset that achieves the goal of
encapsulating formal methods for the railway domain. Overall, the
OnTrack toolset is a modelling and verification environment that
allows graphical scheme plan descriptions to be captured and supported
by formal verification. Thus, it provides a bridge between railway
domain notations and formal specification. This meets the third aim of
this paper, namely to make formal methods accessible to domain
engineers.

In this section we describe the main aspects of the OnTrack tool
including the architecture of the tool. Along with this, we present
the model transformations required to generate \CASL models. This
discussion serves as an illustration on how the tool can be extended
for other formalisms. For example, OnTrack currently also contains the
ability to output models formulated using the \CSPB specification
language~\cite{cspb}. Finally, OnTrack can also generate scheme plan
abstractions, however we refrain from a discussion of these aspects
here and instead refer the reader to~\cite{nfm2013}.

\subsubsection{The OnTrack Toolset Architecture}

OnTrack has been created using the GMF framework~\cite{gronback09} and
multiple Epsilon~\cite{kolovos2012} model transformations.
Figure~\ref{fig:workflow} shows the architecture that we employ in
OnTrack. Initially, a user draws a {\em Track Plan} using the
graphical front end. Then the first transformation, {\em Generate
  Tables} leads to a {\em Scheme Plan}, which is a track plan and its
associated control tables. Generation of control tables has been
previously studied~\cite{mirabadi2009} and here we implement a
technique that produces control tables based on track topology and
signal positions, see~\cite{nfm2013} for details. Track plans and
scheme plans are models formulated relative to our DSL meta model, see
Figure~\ref{fig:bjoerners_dsl_pic}. A scheme plan is then the basis
for subsequent workflows that support its verification. Scheme plans
can then be translated to formal specifications. This can be achieved
in two possible ways, indicated by the optional dashed box in
Figure~\ref{fig:workflow}:

\begin{figure}[h]
\centering
\includegraphics[width=\linewidth]{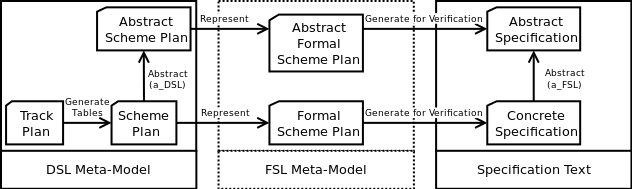}
\caption{The OnTrack workflow.}
\label{fig:workflow}
\end{figure}

\begin{enumerate}
\item Using a meta model for the formal specification language: The
  first option is to have a meta model describing the formal
  specification language. A \emph{Represent} transformation translates
  a \emph{Scheme Plan} into an equivalent \emph{Formal Scheme Plan}
  over the meta model of the formal specification language. Then
  various \emph{Generate for Verification} model to text
  transformations turn a \emph{Formal Scheme Plan} into a \emph{Formal
    Specification Text} ready for verification.
\item Direct generation of a formal specification: The second approach
  is to directly generate a formal specification. Thus only the
  \emph{Generate for Verification} model to text transformations need
  to be implemented.
\end{enumerate}

\noindent In both cases, the \emph{Generate for Verification}
transformations can enrich the models appropriately for verification,
e.g.\ by including the DSL lemmas discussed earlier.

This horizontal workflow provides a transformation yielding a formal
specification that faithfully represents a scheme plan. Here, we
highlight the second approach that has been taken for the generation
of \CASL. The top level of the workflow shows the ability of OnTrack
to include abstractions. We are interested in abstractions to ease
verification. As abstractions do not play a role in our presentation
in this paper, we refer the reader to~\cite{nfm2013} for the details.

\begin{figure}[h]
 \centering
  \includegraphics[width=\linewidth]{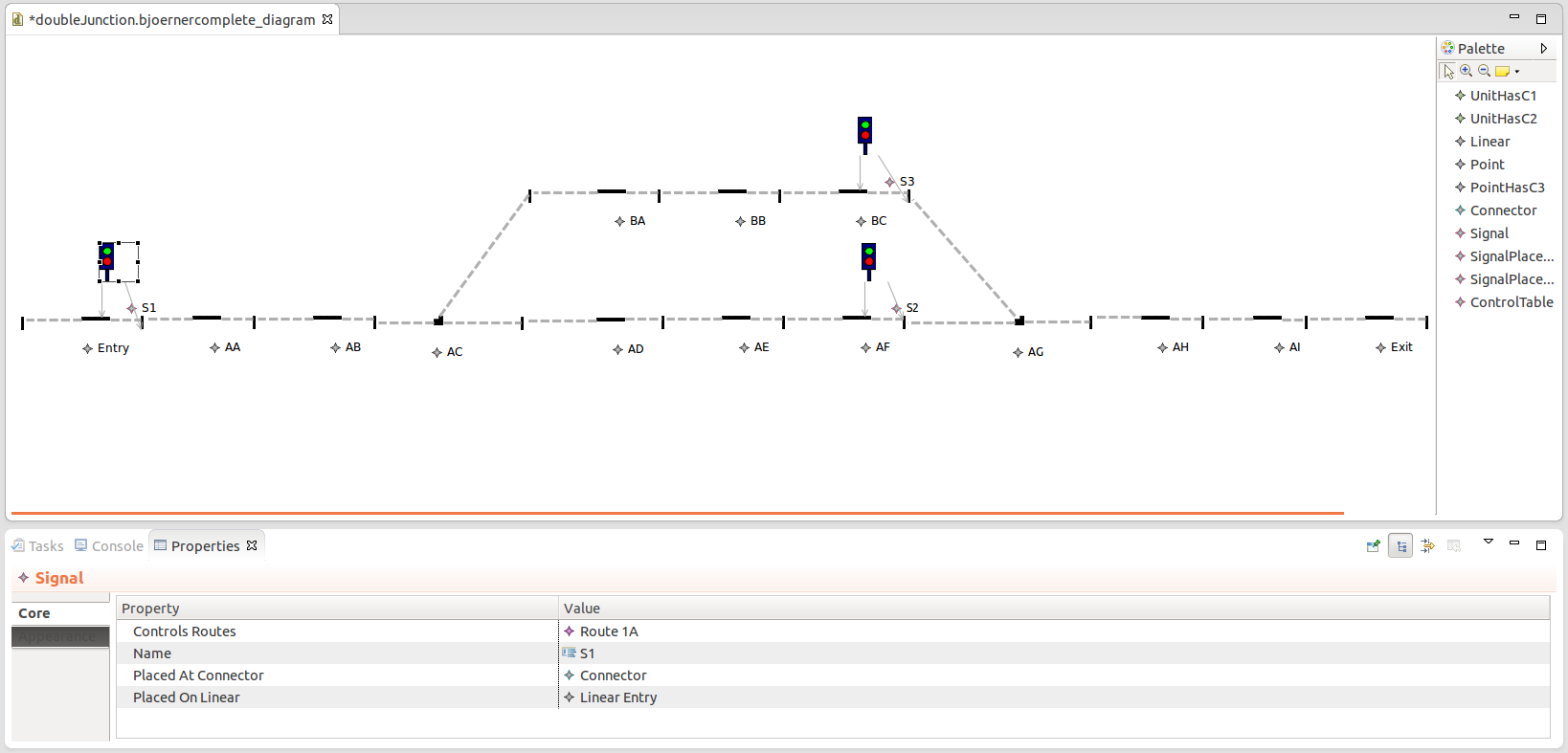}
  \caption{A screenshot of OnTrack modelling a station.}
  \label{fig:editor}
\end{figure}

OnTrack implements this overall workflow in a typical EMF/GMF/Epsilon
architecture~\cite{gronback09,steinberg08,kolovos2012}.  As a basis
for our tool, we have taken the UML diagram in
Figure~\ref{fig:bjoerners_dsl_pic} as our meta-model. Implementing a
GMF front-end for this meta model involves selecting the concepts of
the meta model that should become graphical constructs within the
editor and assigning graphical images to them. Figure~\ref{fig:editor}
shows the OnTrack editor which consists of a drawing canvas and a
palette. Graphical elements from the palette can be positioned onto
the drawing canvas. For example, a linear unit is now a drawable
element.

\subsubsection{Generation of Formal CASL Specifications}

Here we describe the direct implementation of the {\em Generate for
  Verification} transformations for \CASL. The \emph{Generate for
  Verification} transformation translates meta model instances of
\Bjoerner's DSL into formal specification text. This transformation is
implemented using the Epsilon Generation Language
(EGL)~\cite{kolovos2012}. EGL allows template files to be written
describing the text to be generated. These templates provide two main
features for outputting text, namely the ability to output
\emph{static text} and to output \emph{dynamic text}. \emph{Static
  text} is considered text that is always generated independent of the
model. Whilst \emph{Dynamic text} is text that is text dependent on
the given model. For this reason, \emph{Dynamic text} is sometimes
referred to as configuration data. By default, any text written in an
EGL template is considered to be static text. For example we know that
the specification of datatypes for our DSL and similarly our extension
of this with dynamical aspects is the same for all models. Hence this
is rather straightforwardly encoded as static text to be output, see
Figure~\ref{fig:static_gen}.

\begin{figure}[h]
\begin{small}
\begin{framed}
\begin{verbatim}
    spec Pair [sort S] [sort T] =
        sort Pair[S,T]
        ops first: Pair[S,T] -> S;
            second: Pair[S,T] -> T;
    ...
    spec StaticSignature =
         sorts Net, Station, Unit, Connector ...
         sorts Linear, Switch < Unit ...
         preds  __hasLine__: Net * Line;                                
    ...
\end{verbatim}
\vspace{-0.2cm}
\end{framed}
\end{small}
\caption{Example of static text generation for our DSL.}
\label{fig:static_gen}
\end{figure}

Within the same EGL template, we can then specify the output for a
concrete scheme plan. For example, consider the free type of
units. Such a free type is built from the concrete elements of linear
units and switch points contained within the graphical model. Hence we
can specify the template in Figure~\ref{fig:dyn_gen} for the dynamic
generation of the free type Unit. The result of applying this EGL
fragment, to the concrete scheme plan in Figure~\ref{fig:trackplan} is
the following \CASL specification fragment:
\begin{center}
\texttt{free type Unit ::=  lA1 | P1 | PLAT1 | PLAT2 | P2 | LA2}.
\end{center}

\begin{figure}[h]
\begin{small}
\begin{framed}
\begin{verbatim}
1.  [% var rail : RailDiagram  := RailDiagram.allInstances().at(0); %]  
2.  ...
3.  [%if(rail.hasUnits.size > 0){%]
4.      free type Unit ::= 
5.      [%  var i := 0;
6.          while (i < rail.hasUnits.size()-1){ %]
7.          [%  var unit : Unit := rail.hasUnits.at(i);
8.             i := i+1; %]
9.             [%=unit.id%] | [%}%]
10.     [% var unit : Unit := rail.hasUnits.at(i); %]
11.     [%=unit.id%] [%}%]
\end{verbatim}
\vspace{-0.2cm}
\end{framed}
\end{small}
\caption{Dynamic text generation for elements of Unit free type.}
\label{fig:dyn_gen}
\end{figure}

Considering Figure~\ref{fig:dyn_gen}, the first construct of EGL that
we notice is \texttt{[\%} and \texttt{\%]}. Any text specified between
such a set of brackets is interpreted as code. For example, the line
\texttt{var rail : RailDiagram := ...} is a line of EGL code for
declaring the variable \texttt{rail} and assigning to it the current
scheme plan instance within the graphical editor. This variable, can
then be used throughout the EGL template to refer to the current model
instance. Next, we see an EGL if statement (line 3). This statement
checks the number of elements in the \texttt{hasUnits} relation of the
current rail diagram. If there are linear units or points that have
been drawn in the diagram, the code inside the if statement is
executed. The first line within the if statement is static text to be
generated. That is, as long as the if statement is entered, the text
``\texttt{free type Unit ::=}'' will be output. Lines 6 through to 9
perform a loop through the units of the concrete scheme plan
instance. For each unit up until the last but one we can see, that in
line 9, the dynamic text generation ``\texttt{[\%=unit.id\%]}'' is
executed. Here, the dynamic text generation also contains the
``\texttt{=}'' symbol. This indicates that the text following is a
piece of code that returns a value. For example, ``\texttt{unit.id}''
is a field containing the name that has been given to the current unit
element.  This name will then be output by the generation
process. This dynamic text generation is immediately followed by the
static text generation ``\texttt{|}''. This produces the
``\texttt{|}'' symbol between elements of the free type. Finally,
after the while statement there is another block of code (lines 10 and
11) that outputs the last unit identifier in the collection. For this
unit, there is no static generation of the ``\texttt{|}'' symbol which
matches the expected output for a \CASL free type.

In a similar manner to the presented free type generation, it is
possible to explore all the elements of the diagram generating the
concrete scheme plan specification in \CASL. After generation of the
scheme plan, the safety property to be proven over the scheme plan can
also be generated. As this property is the same for all scheme plans,
this is simply generated as static text. The result is a full \CASL
specification ready for verification of the current editor model
instance.

Overall, this generation means that OnTrack achieves the aim of
automating the production of formal specifications from a graphical
model.  OnTrack is a toolset that is usable by engineers from the
railway domain and allows them to produce formal \CASL specifications
ready for verification. This meets the third objective of our
methodology as it makes formal specification available to domain
engineers in an accessible format. The overall result from our
methodology is a graphical tooling environment that incorporates a
``push button'' verification process for the railway signalling
domain.


%% file: 63_verification.tex
\subsection{Contribution: Verification Results}
\label{ssec:verification}
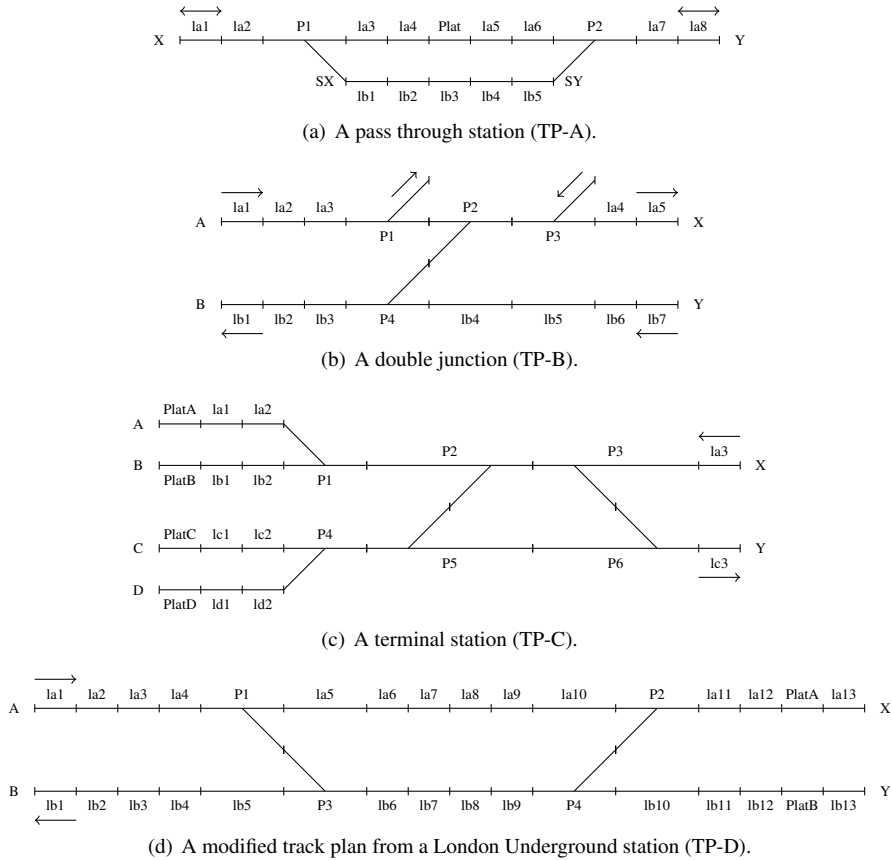
\begin{figure}[h]
    \centering

    \subfigure[A pass through station (TP-A).]
    {       
\begin{tikzpicture}[scale=0.55,transform shape]
\node at (-0.5,0) {X};
\node at (13.5,0) {Y};
\node at (3.5,-1) {SX};
\node at (9.5,-1) {SY};
\draw [<->] (0,0.7) to (1,0.7);
\draw [<->] (13,0.7) to (12,0.7);
\RWConnector{a}{(0,0)}
\RWConnector{a1}{(1,0)}
\RWLinearUnitAbove{a}{a1}{la1}
\RWPointUpsideDown{p1}{p1cl}{p1cn}{p1cr}{(3,0)}
\RWLabelLinearUnitAbove{p1cn}{p1cl}{P1}
\RWLinearUnitAbove{a1}{p1cl}{la2}
\RWConnector{a3}{(5,0)}
\RWLinearUnitAbove{p1cn}{a3}{la3}
\RWConnector{a4}{(6,0)}
\RWLinearUnitAbove{a3}{a4}{la4}
\RWConnector{a5}{(7,0)}
\RWLinearUnitAbove{a4}{a5}{Plat}
\RWConnector{a6}{(8,0)}
\RWLinearUnitAbove{a5}{a6}{la5}
\RWPointReverseUpsideDown{p2}{p2cl}{p2cn}{p2cr}{(10,0)}
\RWLabelLinearUnitAbove{p2cn}{p2cl}{P2}
\RWLinearUnitAbove{a6}{p2cn}{la6}
\RWConnector{a7}{(12,0)}
\RWLinearUnitAbove{p2cl}{a7}{la7}
\RWConnector{a8}{(13,0)}
\RWLinearUnitAbove{a7}{a8}{la8}
\RWConnector{b1}{(5,-1)}
\RWLinearUnitBelow{p1cr}{b1}{lb1}
\RWConnector{b2}{(6,-1)}
\RWLinearUnitBelow{b1}{b2}{lb2}
\RWConnector{b3}{(7,-1)}
\RWLinearUnitBelow{b2}{b3}{lb3}
\RWConnector{b4}{(8,-1)}
\RWLinearUnitBelow{b3}{b4}{lb4}
\RWLinearUnitBelow{b4}{p2cr}{lb5}
\end{tikzpicture}
        \label{fig:first_sub}
    }
    \subfigure[A double junction (TP-B).]
    {
\begin{tikzpicture}[scale=0.55,transform shape]
\node at (-0.5,0) {A};
\node at (-0.5,-2) {B};
\node at (11.5,0) {X};
\node at (11.5,-2) {Y};
\draw [->] (0,0.7) to (1,0.7);
\draw [->] (11,-2.7) to (10,-2.7);
\draw [->] (1,-2.7) to (0,-2.7);
\draw [->] (10,0.7) to (11,0.7);
\RWConnector{a}{(0,0)}
\RWConnector{a1}{(1,0)}
\RWLinearUnitAbove{a}{a1}{la1}
\RWConnector{a2}{(2,0)}
\RWLinearUnitAbove{a1}{a2}{la2}
\RWPoint{p1}{p1cl}{p1cn}{p1cr}{(4,0)}
\RWLabelLinearUnitBelow{p1cn}{p1cl}{P1}
\RWLinearUnitAbove{a2}{p1cl}{la3}
\draw [->] ($(p1) + (0.1,0.6) $) to ($(p1cr) + (-0.3,0.2)$);
\RWPointReverseUpsideDown{p2}{p2cl}{p2cn}{p2cr}{(6,0)}
\RWLabelLinearUnitAbove{p2cn}{p2cl}{P2}
\RWPoint{p3}{p3cl}{p3cn}{p3cr}{(8,0)}
\RWLabelLinearUnitBelow{p3cn}{p3cl}{P3}
\draw [<-] ($(p3) + (0.1,0.6) $) to ($(p3cr) + (-0.3,0.2)$);
\RWConnector{a3}{(10,0)}
\RWLinearUnitAbove{p3cn}{a3}{la4}
\RWConnector{x}{(11,0)}
\RWLinearUnitAbove{a3}{x}{la5}
\RWConnector{b}{(0,-2)}
\RWConnector{b1}{(1,-2)}
\RWLinearUnitBelow{b}{b1}{lb1}
\RWConnector{b2}{(2,-2)}
\RWLinearUnitBelow{b1}{b2}{lb2}
\RWPoint{p4}{p4cl}{p4cn}{p4cr}{(4,-2)}
\RWLabelLinearUnitBelow{p4cn}{p4cl}{P4}
\RWLinearUnitBelow{b2}{p4cl}{lb3}
\RWConnector{b3}{(7,-2)}
\RWLinearUnitBelow{p4cn}{b3}{lb4}
\RWConnector{b4}{(9,-2)}
\RWLinearUnitBelow{b3}{b4}{lb5}
\RWConnector{b5}{(10,-2)}
\RWLinearUnitBelow{b4}{b5}{lb6}
\RWConnector{y}{(11,-2)}
\RWLinearUnitBelow{b5}{y}{lb7}
\end{tikzpicture}     
        \label{fig:second_sub}
    }
    \\
    \subfigure[A terminal station (TP-C).]
    {
\begin{tikzpicture}[scale=0.55,transform shape]
\node at (-0.5,0) {A};
\node at (-0.5,-1) {B};
\node at (-0.5,-3) {C};
\node at (-0.5,-4) {D};
\node at (14.5,-1) {X};
\node at (14.5,-3) {Y};
\draw [->] (14,-0.3) to (13,-0.3);
\draw [->] (13,-3.7) to (14,-3.7);
\RWConnector{a}{(0,0)}
\RWConnector{a1}{(1,0)}
\RWLinearUnitAbove{a}{a1}{PlatA}
\RWConnector{a2}{(2,0)}
\RWLinearUnitAbove{a1}{a2}{la1}
\RWPointReverse{p1}{p1cl}{p1cn}{p1cr}{(4,-1)}
\RWLabelLinearUnitBelow{p1cn}{p1cl}{P1}
\RWLinearUnitAbove{a2}{p1cr}{la2}
\RWLongPointReverseUpsideDown{p2}{p2cn}{p2cl}{p2cr}{(8,-1)}
\RWLabelLinearUnitAbove{p2cn}{p1cl}{P2}
\RWLongPointUpsideDown{p3}{p3cl}{p3cn}{p3cr}{(10,-1)}
\RWLabelLinearUnitAbove{p3cn}{p3cl}{P3}
\RWConnector{a3}{(14,-1)}
\RWLinearUnitAbove{p3cn}{a3}{la3}
\RWConnector{b}{(0,-1)}
\RWConnector{b1}{(1,-1)}
\RWLinearUnitBelow{b}{b1}{PlatB}
\RWConnector{b2}{(2,-1)}
\RWLinearUnitBelow{b1}{b2}{lb1}
\RWLinearUnitBelow{b2}{p1cn}{lb2}
\RWConnector{c}{(0,-3)}
\RWConnector{c1}{(1,-3)}
\RWLinearUnitAbove{c}{c1}{PlatC}
\RWConnector{c2}{(2,-3)}
\RWLinearUnitAbove{c1}{c2}{lc1}
\RWPointReverseUpsideDown{p4}{p4cl}{p4cn}{p4cr}{(4,-3)}
\RWLabelLinearUnitAbove{p4cn}{p4cl}{P4}
\RWLongPoint{p5}{p5cl}{p5cn}{p5cr}{(6,-3)}
\RWLabelLinearUnitBelow{p5cn}{p5cl}{P5}
\RWLinearUnitAbove{c2}{p4cn}{lc2}
\RWLongPointReverse{p6}{p6cl}{p6cn}{p6cr}{(12,-3)}
\RWLabelLinearUnitBelow{p6cn}{p6cl}{P6}
\RWConnector{y}{(14,-3)}
\RWLinearUnitBelow{p6cl}{y}{lc3}
\RWConnector{d}{(0,-4)}
\RWConnector{d1}{(1,-4)}
\RWLinearUnitBelow{d}{d1}{PlatD}
\RWConnector{d2}{(2,-4)}
\RWLinearUnitBelow{d1}{d2}{ld1}
\RWLinearUnitBelow{d2}{p4cr}{ld2}
\end{tikzpicture}
        \label{fig:third_sub}
    }
  \subfigure[A modified track plan from a London Underground
  station (TP-D).]
    {
\begin{tikzpicture}[scale=0.55,transform shape]
\node at (-0.5,0) {A};
\node at (-0.5,-2) {B};
\node at (20.5,-0) {X};
\node at (20.5,-2) {Y};
\draw [->] (0,0.7) to (1,0.7);
\draw [->] (1,-2.7) to (0,-2.7);
\RWConnector{a}{(0,0)}
\RWConnector{a1}{(1,0)}
\RWLinearUnitAbove{a}{a1}{la1}
\RWConnector{a2}{(2,0)}
\RWLinearUnitAbove{a1}{a2}{la2}
\RWConnector{a3}{(3,0)}
\RWLinearUnitAbove{a2}{a3}{la3}
\RWPointUpsideDown{p1}{p1cl}{p1cn}{p1cr}{(5,0)}
\RWLabelLinearUnitAbove{p1cl}{p1cn}{P1}
\RWLinearUnitAbove{a3}{p1cl}{la4}
\RWConnector{a4}{(8,0)}
\RWLinearUnitAbove{p1cn}{a4}{la5}
\RWConnector{a5}{(9,0)}
\RWLinearUnitAbove{a4}{a5}{la6}
\RWConnector{a6}{(10,0)}
\RWLinearUnitAbove{a5}{a6}{la7}
\RWConnector{a7}{(11,0)}
\RWLinearUnitAbove{a6}{a7}{la8}
\RWConnector{a8}{(12,0)}
\RWLinearUnitAbove{a7}{a8}{la9}
\RWPointReverseUpsideDown{p2}{p2cl}{p2cn}{p2cr}{(15,0)}
\RWLabelLinearUnitAbove{p2cl}{p2cn}{P2}
\RWLinearUnitAbove{a8}{p2cn}{la10}
\RWConnector{a9}{(17,0)}
\RWLinearUnitAbove{p2cl}{a9}{la11}
\RWConnector{a10}{(18,0)}
\RWLinearUnitAbove{a9}{a10}{la12}
\RWConnector{a11}{(19,0)}
\RWLinearUnitAbove{a10}{a11}{PlatA}
\RWConnector{a12}{(20,0)}
\RWLinearUnitAbove{a11}{a12}{la13}
\RWConnector{b}{(0,-2)}
\RWConnector{b1}{(1,-2)}
\RWLinearUnitBelow{b}{b1}{lb1}
\RWConnector{b2}{(2,-2)}
\RWLinearUnitBelow{b1}{b2}{lb2}
\RWConnector{b3}{(3,-2)}
\RWLinearUnitBelow{b2}{b3}{lb3}
\RWConnector{b4}{(4,-2)}
\RWLinearUnitBelow{b3}{b4}{lb4}
\RWPointReverse{p3}{p3cl}{p3cn}{p3cr}{(7,-2)}
\RWLabelLinearUnitBelow{p3cl}{p3cn}{P3}
\RWLinearUnitBelow{b4}{p3cn}{lb5}
\RWConnector{b5}{(9,-2)}
\RWLinearUnitBelow{p3cl}{b5}{lb6}
\RWConnector{b6}{(10,-2)}
\RWLinearUnitBelow{b5}{b6}{lb7}
\RWConnector{b7}{(11,-2)}
\RWLinearUnitBelow{b6}{b7}{lb8}
\RWPoint{p4}{p4cl}{p4cn}{p4cr}{(13,-2)}
\RWLabelLinearUnitBelow{p4cl}{p4cn}{P4}
\RWLinearUnitBelow{b7}{p4cl}{lb9}
\RWConnector{b8}{(16,-2)}
\RWLinearUnitBelow{p4cn}{b8}{lb10}
\RWConnector{b9}{(17,-2)}
\RWLinearUnitBelow{b8}{b9}{lb11}
\RWConnector{b10}{(18,-2)}
\RWLinearUnitBelow{b9}{b10}{lb12}
\RWConnector{b11}{(19,-2)}
\RWLinearUnitBelow{b10}{b11}{PlatB}
\RWConnector{b12}{(20,-2)}
\RWLinearUnitBelow{b11}{b12}{lb13}
\end{tikzpicture}
}
    \caption{Verified track plans.}
    \label{fig:plans}
\end{figure}

Each of the track plans in Figure~\ref{fig:plans} (TP-A to TP-D) have
been modelled using our DSL.

We have then included a then \THEN {\small{}\KW{\%}\KW{implies}} block
for the proof of safety. This block is also generated automatically by
the OnTrack tool, as all information required for this block is
available from the graphical model. The block is structured in two
parts, the first contains lemmas to be proven, then the second our
overall proof goal. The aim of the lemmas from the first block is to
encode a case distinction used in the proof of safety. That is, they
help the automated prover in proving our final goal. For verification
using our DSL, these lemmas are in the form of our final proof goal
instantiated for each route of the scheme plan under
consideration. That is, for a route R, we have:
\begin{small}
\begin{hetcasl}
\> \Ax{\forall} \Id{t} \Ax{:} \Id{Time}; \Id{rg} \Ax{:} \IdApplLabel{\Id{Region}}{Region}; \Id{ma} \Ax{:} \Id{MA} \\
\>\> \Ax{\bullet} \=\IdApplLabel{\Id{assigned}}{assigned}(\=\Id{ma}, \Id{t}) \Ax{\wedge} \=\Id{rg} \IdApplLabel{\Id{eps}}{::eps::} \Id{ma} \Ax{\wedge} \=\Id{rg} \IdApplLabel{\Id{eps}}{::eps::} \IdApplLabel{\Id{regions}}{regions}(\Id{R}) 
 \Ax{\Rightarrow} \Ax{\neg} \=\Id{R} \IdApplLabel{\Id{isOpenAt}}{::isOpenAt::} \Id{t} 
\end{hetcasl}
\end{small}

\noindent Once the above implied axioms have been proven, they can be used to
help deduce our overall proof goal:
\begin{small}
\begin{hetcasl}
\THEN \={\small{}\KW{\%}\KW{implies}}\\
\> \Ax{\forall} \Id{t} \Ax{:} \Id{Time}; \Id{r} \Ax{:} \IdApplLabel{\Id{Route}}{Route}; \Id{rg} \Ax{:} \IdApplLabel{\Id{Region}}{Region}; \Id{ma} \Ax{:} \Id{MA} \\
\>\> \Ax{\bullet} \=\IdApplLabel{\Id{assigned}}{assigned}(\=\Id{ma}, \Id{t}) \Ax{\wedge} \=\Id{rg} \IdApplLabel{\Id{eps}}{::eps::} \Id{ma} \Ax{\wedge} \=\Id{rg} \IdApplLabel{\Id{eps}}{::eps::} \IdApplLabel{\Id{regions}}{regions}(\Id{r}) 
\Ax{\Rightarrow} \Ax{\neg} \=\Id{r} \IdApplLabel{\Id{isOpenAt}}{::isOpenAt::} \Id{t} 
\end{hetcasl}
\end{small}

The verification times presented in Figure~\ref{fig:times} show that
verification is possible over our enriched DSL. The average memory
column shows the average memory used across all route lemma proofs and
the safety proof. The proofs have been performed on a quad core $3$GHz
machine with $8$GB of RAM running Ubuntu 12.04.

\begin{figure}[h]
\begin{small}
\centering
\begin{tabular}{|l|r|r|r|}
\hline
Track Plan & Routes Lemma Proofs (s) & Safety Proof (s) & Avg. Memory (MB)\\
\hline  \hline

TP-A &  54.54  & 10.04 & 176.88 \\
TP-B &  23.88  & 6.57 & 90.36 \\
TP-C &  194.41  & 22.34 & 188.07 \\
TP-D &  236.10  & 20.46 & 489.64 \\
\hline  
\end{tabular}
\end{small}
\caption{Verification times for the given track plans.}
\label{fig:times}
\end{figure}

All proofs are completed relatively quickly, with the longest proof
time being for track plan TP-D. Interestingly, this is due to the
number of units contained within this track plan. Also, it is
interesting to note that the track plans that look more complicated
and contain more possible routes, i.e.\ TP-B and TP-C are relatively
quick to verify. This is because these track plans get split into many
smaller regions compared with the fewer larger regions of track plan
TP-D. Each of these small regions is represented by a list containing
fewer elements within our modelling. Hence, the automated theorem
prover does not need to search to such a depth to find a proof. This
point is also illustrated by the increased average memory usage for
TP-D. This shows that the DSL lemmas we have introduced give a
measurable effect for verification. We note that these times
outperform the typical times for model checking on such stations as,
e.g., as presented in \cite{MNRST12HVC}.


%% file: 7_relatedWork.tex
At this point, we reflect on other DSL based approaches for railway
verification, commenting on how they differ to our presented
methodology.

The Railway Control Systems Domain language (RCSD) is a DSL for
railway control systems developed by Kirsten
Mewes~\cite{mewes2010}. RCSD is motivated by model-driven engineering
approaches to system design. The language uses common domain notation
for its concrete syntax and incorporates knowledge from domain
engineers about the domain into its static and dynamic
semantics. Mewes also considers the topic of testing models created in
DSL. Here, domain specific constraints are included into the models
ensuring that validation checks for correct functionality can be
tested at the model level, before any software is been
developed. Mewes' work focuses on the design of RCSD. In contrast, we
leave the design of the DSL to the domain engineer and focus on
capturing domain knowledge that can be exploited for automatic
verification.

Work by Haxthausen and Peleska~\cite{peleska07} has also explored the
development of a domain specific framework for automated construction
and verification of railway control systems. Their framework consists
of a three tiered approach: the top layer is a DSL for use by domain
engineers to specify railway control systems; the second tier is model
generation: A generator automatically produces the model of a control
program based on the specification given by the design
engineer. Bounded model checking can then be performed to establish
various safety properties over such programs; the third tier allows
actual code to be automatically generated and verified to ensure
certain properties are maintained throughout process. Differing to our
work, this framework is specifically developed for the railway domain
and tied to a specific DSL. In this paper, we provide a generic and
systematic methodology, where the railway domain serves for
illustration.

Finally, the SafeCap toolset~\cite{iliasov2012b} provides a tooling
platform that supports reasoning about railway capacity whilst
ensuring system safety. It is based upon the SafeCap DSL, which
captures track topology, route and path definitions and signalling
rules. Overall, the toolset allows signalling engineers to design
stations and junctions, to check their safety and to evaluate the
potential improvements in capacity. Again, the SafeCap approach is
bound to a specific DSL, where the design of the DSL is tailored to
the underlying verification technology (Event B). This differs from
our aim to decouple the DSL from the formal specification language, in
turn allowing tooling environments to be very openly extendable.


%% file: 8_reflection.tex
Since 2007, the Railway Verification Group at Swansea University
Computer Science has been working in collaboration with Invensys
Rail. The group is supported by eight academics. It adopts formal
techniques to railway systems. This setup has led to a rich process of
information exchange, by mutual visits and internships, as well as
regular meetings. Out of this collaboration, there have been several
successful research projects covering research into verification of
interlocking
programs~\cite{kanso08b,kansoThesis,james10b,james10a,lawrence11,james13b},
the verification of scheme plan
designs~\cite{MNRST12,MNRST12HVC,nfm2013,james13a}, and capacity
analysis~\cite{isobeMNR12}. These projects have involved many
specification formalisms including propositional logic, CSP, Timed
CSP, \CSPB, Scade, \CASL and Agda. Here, it appears that operational
based models such as the one we have provided in \CASL are easier for
railway engineers to follow and understand.

The methodology outlined in this paper has also been successfully
applied to the Invensys Rail Data Model~\cite{DataModel},
see~\cite{james14} for full details. In this context, the DSL
definition is provided by Invensys Rail in the form of UML class
diagrams with accompanying narrative. This DSL definition has been
completely designed by Invensys Rail without our support. Hence,
developing the informal DSL that is used as a starting point for our
methodology is clearly a task that can be undertaken purely within
industry. With regards to the formalisation of the DSL in \CASL, this
step was performed in a cyclic manner. Namely, the computer scientists
would suggest a formalisation at a meeting with Invensys Rail, and
then feedback would be provided on the formalisation. This process was
repeated until both groups were happy with the resulting
formalisation. Verification support (in terms of supporting lemmas)
was then developed purely by the computer scientists at Swansea. As
this step does not alter the DSL, the engineers were happy with the
approach.

Finally, the graphical front end to the OnTrack tool provides a view
to scheme plans that is of the same nature as representations used
within Invensys Rail. The railway engineers have seen the verification
process we propose and are confident that they could follow it thanks
to the automated nature of the process. They have expressed an
interest in the extension of OnTrack to allow for importing and
exporting of models using an industrial format such as the XLDL (XML
Layout Description Language) format used by Invensys Rail. This would
allow the toolset to be more easily integrated within current
development processes at Invensys Rail. We note that even though
Invensys Rail believe technologies such as the OnTrack toolset will
improve their development processes, due to concerns around the
integrity of automated verification tools, quality assurance of a
scheme plan design will, in the near future, still rely upon
traditional testing and inspection. Here, tool qualification is the
issue: our tools are not yet ``proven-in-use'', i.e., there is no
experience from previous industrial safety-critical projects
suggesting that our tools are ``correct''; alternatively, our tools
are not yet certified by a designated authority.

Finally, it remains future work to perform a systematic evaluation
into how usable the presented verification is for railway engineers.
To do this, one could carry out a pilot project. This would involve
thorough time measurement and documentation for all activities that
our approach incurs including installing, using, and integrating
OnTrack into the development process. Reflecting upon these results
would allow us to check if our approach is indeed feasible. On the
management level, such a pilot project would then have to be evaluated
as to whether it is an improvement with respect to current
practice. It should be both more effective, in that it allows one to
achieve better results than current practice, and more efficient,
i.e., it should offer a better cost/result ratio.


%% file: 9_conclusion.tex
In this paper, we have introduced a novel design methodology for
encapsulating formal methods within DSLs. We have supported our
hypothesis that DSLs can aid with verification by showing it to be
valid for the railway domain. 

Our methodology begins with industrial documents describing domain
elements in the form of UML class diagrams with narratives. This
informal DSL is formalised in the algebraic specification
language~\CASL, where we support this step with automated translations
of the UML class diagram, thus ensuring ``faithful modelling'' of the
domain. Next, systematic exploitation of domain knowledge allows us to
prove domain specific lemmas. Thanks to these domain specific lemmas,
properties on the formal specifications can automatically be proven
for a class of systems, thus addressing issues surrounding
``scalability''. Finally, a graphical editor for the DSL is
developed. This editor allows engineers from the domain to follow a
seamless verification process: (1) they can formulate models in the
DSL; (2) these models can be automatically transformed into formal
specifications; and (3) they can automatically verify these models
thanks to the domain specific lemmas. Overall, this addresses the
issues of ``accessibility''of formal methods.

The verification of railway control software has been identified as a
grand challenge of computer science~\cite{topic11}. The defining
element of a grand challenge is that progress towards the challenge
results in progress in computer science in general. Therefore,
motivated by the railway domain, our methodology can be seen as a step
forward for industrial applications of formal methods.

Along with presenting this methodology, we have also successfully
illustrated the use of algebraic specification for modelling railway
systems and the use of automated theorem proving for railway
verification. Bringing these points together results in a strong case
for using DSLs in the setting of specification and verification.

In the future, we would like to explore visual feedback of failed
proof attempts. Currently, feedback to the user is provided in the
form of a named route which is unsafe (obtained from the name of the
failed proof). Such visualisations have been considered by Marchi et
al.\ \cite{ic2011}, and it would be interesting to extend OnTrack with
such visualisations. We would also like to illustrate the
applicability of our methodology to further domains, such as to the
design of medical devices as considered by Oladimeji et
al.\ \cite{olad13}.
\\
\paragraph{Acknowledgements:}
The authors would like to thank Simon Chadwick and Dominic Taylor from
Invensys Rail UK for their contributions and encouraging feedback. We
would like to thank our colleagues from the Swansea railway
verification group and the Swansea Processes and Data research group
for their input and feedback towards this work. Similarly, we greatly
appreciate the input of Alexander Knapp and Till Mossakowski towards
our work on the UML institution and comorphism to \ModalCASL. We also
thank Helen Treharne, Steve Schneider and Matthew Trumble from Surrey
University for their collaboration in developing the OnTrack tool. Our
final thanks goes to Erwin R.\ Catesbeiana (Jr) for signalling us in
the correct direction.


%% file: appendix_proof.tex
\section{Domain Specific Lemma Proof}
\label{app:proof}
\textbf{Proof:}
Following from the axioms presented in
Section~\ref{sec:narrative_model}. we have:
  \begin{description}
  \item $\Leftarrow$: Let us assume (**). The proof is given by
    induction on time $t$.  

    In the base case (t=$0$), (*) is true. This is the case as only
    the empty movement authority is assigned at time $0$ (given by
    axiom {\small{}\KW{\%}(no\_ma\_0)\KW{\%}}). Let $m1$, $m2$ be two
    movement authorities such that $share(m1,m2)$ holds. We consider
    two cases: 

    (1) if $m1=m2$ then the implication holds trivially;

    (2) if $m1 \not= m2$ then only one of these movement authorities
    can be the empty movement authority. Let, without loss of
    generality, $m1 \not=[]$. Then $assigned(m1,0)$ is false (by axiom
    {\small{}\KW{\%}(no\_ma\_0)\KW{\%}}), hence the implication holds.

    For the step case we have to show that:
\begin{hetcasl}
\>\Ax{\forall} \Id{t} \Ax{:} \Id{Time} \Ax{\bullet}\\
\>\> (\Ax{\forall} \Id{m1}, \Id{m2} \Ax{:} \Id{MA} \Ax{\bullet} \IdApplLabel{\Id{share}}{share}(\Id{m1}, \Id{m2}) \\
\>\>\> \Ax{\Rightarrow} \Id{m1} \Ax{=} \Id{m2} \Ax{\vee} \Ax{\neg} (\IdApplLabel{\Id{assigned}}{assigned}(\Id{m1}, \IdApplLabel{\Ax{t}}{t}) \Ax{\wedge} \IdApplLabel{\Id{assigned}}{assigned}(\Id{m2}, \IdApplLabel{\Ax{t}}{t}))) {\small{}\KW{\%}(ih)\KW{\%}}  \\
\>\>$\Rightarrow$\\
\>\> (\Ax{\forall} \Id{m1}, \Id{m2} \Ax{:} \Id{MA} \Ax{\bullet} \IdApplLabel{\Id{share}}{share}(\Id{m1}, \Id{m2}) \\
\>\>\> \Ax{\Rightarrow} \Id{m1} \Ax{=} \Id{m2} \Ax{\vee} \Ax{\neg} (\IdApplLabel{\Id{assigned}}{assigned}(\Id{m1}, \IdApplLabel{\Ax{suc(t)}}{suc(t)}) \Ax{\wedge} \IdApplLabel{\Id{assigned}}{assigned}(\Id{m2}, \IdApplLabel{\Ax{suc(t)}}{suc(t)}))) 
\end{hetcasl}

Let us assume (ih) for all movement authorities $m1,m2$. Let $m1'$,
$m2'$ be two movement authorities such that $share(m1',m')$ holds. We
consider two cases: 

(1) if $m1'=m2'$ then the implication holds trivially; 

(2) if $m1' \not= m2'$ then we need to show \Ax{\neg}
(\IdApplLabel{\Id{assigned}}{assigned}(\Id{m1'}, \Id{t}) \Ax{\wedge}
\IdApplLabel{\Id{assigned}}{assigned}(\Id{m2'}, \Id{t})). By case
distinction over the definition of the predicate \Id{assigned} (axiom
{\small{}\KW{\%}(assigned\_defn)\KW{\%}}), we show that
(\IdApplLabel{\Id{assigned}}{assigned}(\Id{m1'}, \Id{t}) \Ax{\wedge}
\IdApplLabel{\Id{assigned}}{assigned}(\Id{m2'}, \Id{t})) is not
possible and hence the implication holds.
\begin{itemize}
\item Case 1: (\IdApplLabel{\Id{assigned}}{assigned}(\Id{m1'},
  \IdApplLabel{\Ax{suc(t)}}{suc(t)}) \Ax{\wedge}
  \IdApplLabel{\Id{assigned}}{assigned}(\Id{m2'},
  \IdApplLabel{\Ax{suc(t)}}{suc(t)})) $\Rightarrow$
  (\IdApplLabel{\Id{assigned}}{assigned}(\Id{m1'}, \Id{t}) \Ax{\wedge}
  \IdApplLabel{\Id{assigned}}{assigned}(\Id{m2'}, \Id{t}) does not
  hold as it contradicts the assumed induction hypothesis (ih).

\item Case 2: (\IdApplLabel{\Id{assigned}}{assigned}(\Id{m1'},
  \IdApplLabel{\Ax{suc(t)}}{suc(t)}) \Ax{\wedge}
  \IdApplLabel{\Id{assigned}}{assigned}(\Id{m2'},
  \IdApplLabel{\Ax{suc(t)}}{suc(t)})) $\Rightarrow$
  \IdApplLabel{\Id{canExtend}}{canExtend}(\Id{m1'}, \Id{t}) \Ax{\wedge}
  \IdApplLabel{\Id{canExtend}}{canExtend}(\Id{m2'}, \Id{t}), does not hold
  as it contradicts with axiom\break
  {\small{}\KW{\%}(one\_ma\_changes)\KW{\%}} stating that only one
  movement authority extends per time step.

\item Case 3: (\IdApplLabel{\Id{assigned}}{assigned}(\Id{m1'},
  \IdApplLabel{\Ax{suc(t)}}{suc(t)}) \Ax{\wedge}
  \IdApplLabel{\Id{assigned}}{assigned}(\Id{m2'},
  \IdApplLabel{\Ax{suc(t)}}{suc(t)})) $\Rightarrow$
  \IdApplLabel{\Id{assigned}}{assigned}(\Id{m1'}, \Id{t}) \Ax{\wedge}
  \IdApplLabel{\Id{canExtend}}{canExtend}(\Id{m2'}, \Id{t}). Given that
  \IdApplLabel{\Id{share}}{share}(\Id{m1'}, \Id{m2'}) holds, by the
  definition of the predicate $share$ (axiom
  {\small{}\KW{\%}(share\_defn)\KW{\%}}) we know that there exists a
  region $rg$ such that $rg \in$ \Id{m1'} and $rg \in$ \Id{m2'}. Let
  us consider, what it means for
  \IdApplLabel{\Id{canExtend}}{canExtend}(\Id{m2'}, \Id{t}) to hold,
  i.e. instantiating axiom {\small{}\KW{\%}(extend\_defn)\KW{\%}} with
  $m2'$ gives:
\begin{hetcasl}
  \IdApplLabel{\Id{canExtend}}{canExtend}(\=\Id{m2'}, \Id{t})
  \Ax{\Leftrightarrow}\\
  \> \=\Ax{\exists} \Id{ma} \Ax{:} \IdApplLabel{\Id{MA}}{MA}; \Id{r} \Ax{:} \IdApplLabel{\Id{Route}}{Route} \Ax{\bullet}\\
  \>\>\> \=\IdApplLabel{\Id{assigned}}{assigned}(\=\Id{ma}, \Id{t})
  \Ax{\wedge} \IdApplLabel{\Id{ext}}{ext}(\=\Id{ma}, \Id{r}, \Id{m2'})
  \Ax{\wedge}
  \=\Id{r} \IdApplLabel{\Id{isOpenAt}}{::isOpenAt::} \Id{t} \Ax{\wedge} \\
  \>\>\> (\=\Ax{\neg} \=\Id{ma} \Ax{=} \=\Ax{[}\Ax{]} \Ax{\Rightarrow}
  \Ax{\neg} \IdApplLabel{\Id{assigned}}{assigned}(\=\Id{ma},
  \IdApplLabel{\Id{suc}}{suc}(\Id{t})))
\end{hetcasl}
Hence, we have two cases to consider for the shared region $rg$: 

(1) if $rg \in$ \Id{ma}, then by (**), all routes $r'$
such that $rg \in regions(r')$ are not open. Hence given axiom
{\small{}\KW{\%}(ext\_defn)\KW{\%}} which tells us that ext acts like
list concatenation we know that $\neg \; r \; isOpenAt \; t$. Thus we
have a contradiction that $canExtend(m2',t)$ must hold, but also $\neg
\; r \; isOpenAt \; t$ must hold; 

(2) if $rg \not\in$ \Id{ma} then we know due to axiom
{\small{}\KW{\%}(ext\_defn)\KW{\%}} that $share(m1',ma)$ holds. We
also know, by the definition of canExtend (axiom
{\small{}\KW{\%}(extends\_defn)\KW{\%}}) that $assigned(ma,t)$ holds
and similarly by our case assumption that \break $assigned(m1',t)$
holds. Now if $m1' \not=ma$ then we gain a contradiction to our
induction hypothesis (ih). Whereas, if $m1' = ma$, then by the
definition of $canExtend$ (axiom
{\small{}\KW{\%}(extends\_defn)\KW{\%}}) we know that $\neg
assigned(ma,suc(t))$ and hence that \break $\neg
assigned(m1',suc(t))$. This again contradicts our case assumption that
$assigned(m1',t)$ holds.

 \item Case 4:
 (\IdApplLabel{\Id{assigned}}{assigned}(\Id{m1'},
  \IdApplLabel{\Ax{suc(t)}}{suc(t)}) \Ax{\wedge}
  \IdApplLabel{\Id{assigned}}{assigned}(\Id{m2'},
  \IdApplLabel{\Ax{suc(t)}}{suc(t)})) $\Rightarrow$
  \IdApplLabel{\Id{canExtend}}{canExtend}(\Id{m1'}, \Id{t})
  \Ax{\wedge} \IdApplLabel{\Id{assigned}}{assigned}(\Id{m2'}, \Id{t}):
  Analogous to Case 3.

\item Case 5-9: The remaining cases involve the reduction of one
  movement authority that is already assigned. Hence, each case
  follows from the definition of the
  \IdDeclLabel{\Id{canReduce}}{canReduce} predicate
  {\small{}\KW{\%}(reduces\_defn)\KW{\%}} and our induction hypothesis
  (ih).
\end{itemize}

\item $\Rightarrow$: Let us assume (*). The proof is by contradiction.

 We assume (**) does not hold, that is, we know there exists a route $r$ such that:
\begin{hetcasl}
\> \Ax{\exists} \Id{t} \Ax{:} \Id{Time}; \Id{r} \Ax{:} \Id{Route}; \Id{rg} \Ax{:} \IdApplLabel{\Id{Region}}{Region}; \Id{ma} \Ax{:} \Id{MA} 
\\ 
\>\>\Ax{\bullet} \IdApplLabel{\Id{assigned}}{assigned}(\Id{ma}, \Id{t}) \Ax{\wedge} \Id{rg} \IdApplLabel{\Id{eps}}{::eps::} \Id{ma} \Ax{\wedge} \Id{rg} \IdApplLabel{\Id{eps}}{::eps::} \IdApplLabel{\Id{regions}}{regions}(\Id{r})\\
\>\>\> \Ax{\Rightarrow} \Id{r} \IdApplLabel{\Id{isOpenAt}}{::isOpenAt::} \Id{t} 
\end{hetcasl}
Given this, this route can be used for an extension to another
movement authority even though it is assigned. Assume at some time
$t'$, such that $suc(t') = t$ there exist movement authorities $ma1,
ma2$ such that $ext(ma1,r ma2)$ and $rg \; eps \; ma2$. Then the predicate
$canExtend$ (axiom {\small{}\KW{\%}(extends\_defn)\KW{\%}}) holds for
$ma2$ at time $t$. Thus, by the definition of assigned (axiom
{\small{}\KW{\%}(assigned\_defn)\KW{\%}}) $ma2$ can become assigned at
time $suc(t')$, i.e.\ time $t$. Now we have the situation that
$assigned(ma,t)$ and $assigned(ma2,t)$ hold, but so does
$share(ma,ma2)$ (axiom {\small{}\KW{\%}(share\_defn)\KW{\%}}) over the
region $rg$. Hence we have a contradiction to (**).
\end{description}

%% file: appendix_spec_structure.tex
\section{Specification Structure}
\label{app:struture}
Throughout the paper, we have introduced various levels of
specification that are structured as follows:

\begin{small}
\begin{hetcasl}
\>\SPEC \SId{DSLForVerification} \Ax{=} \\
\>\> \SId{Datatypes} \\
\>\THEN \\
\>\> \SId{DSL}\\
\> \THEN\\
\>\> \SId{DSLExtension}\\
\>\THEN \={\small{}\KW{\%}\KW{implies}}\\
\>\>  \SId{DSLemmas} \\
\>\THEN \\
\>\> \SId{ConcreteSchemePlan}\\
\>\THEN \={\small{}\KW{\%}\KW{implies}}\\
\>\> \SId{Safety}\\
\>\KW{end}
\end{hetcasl}
\end{small}

\noindent Here we can see, that our specifications begin with the Datatypes
(\SId{Datatypes}) and DSL (\SId{DSL}) gained from our translation of
the UML class diagram. We then extend these specifications with
\SId{DSLExtension} for modelling the narrative aspects of the original
informal DSL. Concretely this includes our modelling of movement
authorities. Next, we can see that the property supporting lemmas
(\SId{DSLemmas}) are added to aid with verification. These are added
as implied axioms over the extended DSL. This illustrates that these
lemmas are independent of any scheme plan formulated in the DSL. Next,
we see the specification \SId{ConcreteSchemePlan} that encodes a
particular track plan and its associated movement authorities and
control tables. Finally, we can see the proof goals for proving safety
are added (\SId{Safety}). These are again added as implied axioms, and
are then proven relative to the given concrete scheme plan.
